\documentclass[aps,showpacs,preprint,superscriptaddress]{revtex4-1}

\usepackage{amsmath,txfonts,longtable,booktabs,appendix,overpic,amssymb,bm,bbm,multirow,float,graphicx,color,dcolumn,subfigure,enumitem}

\usepackage[colorlinks=true,
linkcolor=blue,
citecolor=blue,
urlcolor=blue,
filecolor=blue,
anchorcolor=blue,
menucolor=blue]{hyperref}

\begin{document}

	\title{Spin-resolved momentum distributions and vortex evolution in multiphoton pair production within a modified two-level model}
	
	\author{Qi-Rong Wang}
	\affiliation{Key Laboratory of Beam Technology of the Ministry of Education, and School of Physics and Astronomy, Beijing Normal University, Beijing 100875, China}
	\author{Hong-Hao Fan}
	\affiliation{School of Mathematics and Physics, Xinjiang Hetian College, Hetian 848000, China}
	\author{Musajan Kadir}
	\affiliation{Key Laboratory of Beam Technology of the Ministry of Education, and School of Physics and Astronomy, Beijing Normal University, Beijing 100875, China}
	\author{Ying Wu}
	\affiliation{Key Laboratory of Beam Technology of the Ministry of Education, and School of Physics and Astronomy, Beijing Normal University, Beijing 100875, China}
	\author{Bai-Song Xie}\thanks{\href{mailto:bsxie@bnu.edu.cn}{bsxie@bnu.edu.cn}}
	\affiliation{Key Laboratory of Beam Technology of the Ministry of Education, and School of Physics and Astronomy, Beijing Normal University, Beijing 100875, China}
	\affiliation{Institute of Radiation Technology, Beijing Academy of Science and Technology, Beijing 100875, China}
	
	\date{\today}

	\begin{abstract}
		We investigate spin-resolved momentum distributions and phase vortices in multiphoton electron-positron pair production within a modified two-level model. This model retains the diagonal spinor inner-product terms neglected in the adiabatic approximation, thereby inducing spin-resolved transverse asymmetries in the momentum distribution. For a circularly polarized pulse, the handedness determines the winding direction of the quantized vortices, whereas for a linearly polarized pulse, the phase distribution exhibits nodal lines rather than quantized vortices. For two time-delayed circularly polarized pulses, the time delay introduces a relative phase that governs the multiphoton interference, which in turn drives the vortex evolution in each spin configuration. These results demonstrate that pulse polarization and time delay enable coherent control over vortex topology in pair production.
	\end{abstract}

	\maketitle

	\section{Introduction}

	In the presence of a strong external electric field, the vacuum of quantum electrodynamics (QED) becomes unstable and decays into electron-positron pairs, a phenomenon known as the Sauter-Schwinger effect~\cite{1,2,3,4,5,6}. Two distinct pair-production mechanisms are generally recognized: the nonperturbative tunneling mechanism for \(\gamma \ll 1\)~\cite{7,8}, and the perturbative multiphoton mechanism for \(\gamma \gg 1\)~\cite{9}. Here, \(\gamma = m\omega /(|e|E_0)\) is the Keldysh adiabaticity parameter, where \(m\) and \(e = - |e|\) denote the electron mass and charge, and \(\omega\) and \(E_0\) denote the frequency and amplitude of the external electric field, respectively. Recent investigations of these mechanisms have revealed diverse phenomena, including momentum spectra~\cite{10,11,12,13,14,15,16,17,18,19,20,21,22,23,24}, quantum entanglement~\cite{25,26,27,28}, and vortex structures~\cite{29,30,31,32}.
	
	Vortices are ubiquitous topological structures manifesting across branches of physics, including hydrodynamic flows~\cite{33}, atomic and molecular ionization~\cite{34,35}, electron vortex beams~\cite{36}, and laser-driven quantum dynamics~\cite{37}. In strong-field physics, phase vortices have been identified in the probability amplitudes for photoionization and photodetachment~\cite{37,38,39,40}. More recently, they have been predicted in the probability amplitude for multiphoton pair production~\cite{41,42}. In scalar QED, the topological charge of the vortex is exactly the intrinsic orbital angular momentum (IOAM) of the produced pair rather than that of individual particles~\cite{43}. In spinor QED, by contrast, it is jointly determined by the number of absorbed photons and the spin projection of the pair, providing a direct topological signature of angular momentum transfer from the field to the pair~\cite{44}. Although the topological transitions of vortices in two-colored fields have been explored~\cite{45}, a systematic understanding of vortex formation, evolution, and annihilation in strong fields remains incomplete, which motivates the present work.
	
	In QED pair production, vortex structures have been primarily investigated via the scattering matrix approach~\cite{46,47} and the two-level model (TLM)~\cite{48,49}. The TLM rests on the adiabatic approximation, which neglects the inner products between the time derivatives of the instantaneous eigenvectors and the eigenvectors themselves~\cite{49}. To accurately describe pair production, we employ a modified two-level model (mTLM) that retains these terms, and thereby reveal the transverse asymmetry along the \(p_x\) direction in the spin-resolved momentum distribution.
	
	Based on the mTLM, we investigate the spin-resolved momentum and phase distributions in multiphoton electron-positron pair production. In particular, we focus on the vortex evolution controlled by the time delay between two circularly polarized pulses, with circularly polarized and linearly polarized single pulses serving as reference cases for vortex formation and annihilation, respectively. In addition, we introduce an interference degree to quantify the quantum interference between the two pulses. This work may provide insight into the vortex dynamics in multiphoton pair production.
	
	This paper is organized as follows. Section~\ref{Sec:II} presents the theoretical model for calculating spin-resolved probability amplitudes for electron-positron pair production in a spatially homogeneous, time-dependent electric field. Section~\ref{Sec:III} presents the numerical results. We first investigate the spin-resolved momentum and phase distributions for a single elliptically polarized pulse in Sec.~\ref{Sec:IIIA}, then extend the analysis to two time-delayed circularly polarized pulses to explore the vortex evolution and the interference degree in Sec.~\ref{SecIII:B}, and finally analyze the transverse asymmetry by comparing the TLM and the mTLM in Sec.~\ref{SecIII:C}. Section~\ref{Sec:IV} summarizes the conclusions. Appendix~\ref{App:A} establishes the theoretical foundation by deriving the mTLM evolution equations from the Dirac equation. Appendix~\ref{App:B} analytically derives the time-delay phase factor via the temporal translation. Appendix~\ref{App:C} derives the ratio of the momentum distribution functions under the momentum reflection transformation. We adopt natural units with \(\hbar = c = m = 1\) and set \(e = -1\), with metric signature \((+, -, -, -)\).

	\section{THEORETICAL MODEL}\label{Sec:II}
	
	In this section, we employ the mTLM to investigate electron-positron pair production in a spatially homogeneous, time-dependent electric field \(\bm{E}(t) = (E_x(t),E_y(t),E_z(t))\), with the corresponding vector potential \(\bm{A}(t) = (A_x(t),A_y(t),A_z(t))\) in the temporal gauge. The electric field and the vector potential are related by \(\bm{E}(t) = -\dot{\bm{A}}(t)\).
	
	The Bogoliubov transformation relates the time-independent electron annihilation operators \(a_{\bm{p},s}\) and positron creation operators \(b_{-\bm{p},s'}^{\dagger}\) in the free vacuum to the time-dependent operators \(\tilde{a}_{\bm{p},s}(t)\) and \(\tilde{b}_{-\bm{p},s'}^{\dagger}(t)\) in the instantaneous eigenbasis as follows:
	\begin{equation}\label{eq:1}
		\left( \begin{array}{c}\tilde{a}_{\bm{p},s}(t)\\ \tilde{b}_{-\bm{p},s'}^{\dagger}(t) \end{array} \right) = \left( \begin{array}{cc}\alpha_{s,s'}(\bm{p},t) & -\beta_{s,s'}^{\ast}(\bm{p},t)\\ \beta_{s,s'}(\bm{p},t) & \alpha_{s,s'}^{\ast}(\bm{p},t) \end{array} \right)\left( \begin{array}{c}a_{\bm{p},s}\\ b_{-\bm{p},s'}^{\dagger} \end{array} \right),
	\end{equation}
	where \(\alpha_{s,s'}(\bm{p},t)\) and \(\beta_{s,s'}(\bm{p},t)\) are the Bogoliubov coefficients. Fermi statistics imposes the unitarity condition \(|\alpha_{s,s'}(\bm{p},t)|^{2} + |\beta_{s,s'}(\bm{p},t)|^{2} = 1\), with \(|\beta_{s,s'}(\bm{p},t)|^{2}\) as the momentum distribution for producing an electron-positron pair in the spin configuration \((s,s')\). Here, \(s, s' = \pm 1\) denote the spin projections of the electron and positron along the \(z\)-axis, respectively, with the total spin projection defined as \(\mathcal{S}_{s,s'} = (s + s')/2\). For the four spin configurations, \(\mathcal{S}_{+,+'} = 1\), \(\mathcal{S}_{+,-'} = 0\), \(\mathcal{S}_{-,+'} = 0\) and \(\mathcal{S}_{-,-'} = -1\).                                                                                                                                                                                                                                                                                                                                                                                                                                                                                                                                                                                                                                                                                                                                                                                                                                                                                                                                                                                                                                                                                                                                                                                                                                                                                                                                                                                                                                                                                                                                                                                                                                                                                                                                                                                                                                                                                                                                                                                                                                                       
	
	Introducing the phase-rotated coefficients \(c_{s,s'}^{\alpha}(\bm{p},t) = \alpha_{s,s'}(\bm{p},t)\mathrm{e}^{-\mathrm{i}\int_{-\infty}^{t}\omega_{\bm{p}}(\tau)\mathrm{d}\tau}\) and \(c_{s,s'}^{\beta}(\bm{p},t) = \beta_{s,s'}(\bm{p},t)\mathrm{e}^{+\mathrm{i}\int_{-\infty}^{t}\omega_{\bm{p}}(\tau)\mathrm{d}\tau}\), we obtain the spin-resolved fermionic evolution equation of the mTLM:
	\begin{equation}\label{eq:2}
		\mathrm{i} \begin{pmatrix} \dot{c}_{s,s'}^{\alpha}(\bm{p},t) \\ \dot{c}_{s,s'}^{\beta}(\bm{p},t) \end{pmatrix} = \begin{pmatrix} \omega_{\bm{p}}(t) + \Delta_{s,s'}(\bm{p},t) & \mathrm{i}\Omega_{s,s'}(\bm{p},t) \\ -\mathrm{i}\Omega_{s,s'}^{*}(\bm{p},t) & -\omega_{\bm{p}}(t) + \Delta_{-s,-s'}(\bm{p},t) \end{pmatrix} \begin{pmatrix} c_{s,s'}^{\alpha}(\bm{p},t) \\ c_{s,s'}^{\beta}(\bm{p},t) \end{pmatrix},
	\end{equation}
	where \(\omega_{\bm{p}}(t)\) is the instantaneous energy eigenvalue, and \(\Delta_{s,s'}(\bm{p},t)\), \(\Omega_{s,s'}(\bm{p},t)\) are the spinor inner products. The derivation of Eq.~\eqref{eq:2} from the Dirac equation is given in Appendix~\ref{App:A}. The previous TLM neglects the diagonal spinor inner-product terms \(\Delta_{s,s'}(\boldsymbol{p},t)\) due to the slowly varying envelope approximation~\cite{49}; however, the present mTLM retains these terms.
	
	With the initial conditions \(c_{s,s'}^{\alpha}(\bm{p},-\infty) = 1\) and \(c_{s,s'}^{\beta}(\bm{p},-\infty) = 0\), solving Eq.~\eqref{eq:2} yields the final probability amplitude \(c_{s,s'}^{\beta}(\bm{p},+\infty)\). The electron spin-resolved momentum distribution is then given by
	\begin{equation}\label{eq:3}
		f_{s,s'}(\bm{p}) = |c_{s,s'}^{\beta}(\bm{p},+\infty)|^{2}.
	\end{equation}
	
	We define the phase of the probability amplitude as \(\phi_{s,s'}(\bm{p}) = \arg [c_{s,s'}^{\beta}(\bm{p},+\infty)]\). The phase singularities coincide with isolated zeros of the complex probability amplitude \(c_{s,s'}^{\beta}(\bm{p},+\infty)\), arising from destructive interference among multiphoton pathways. The topological charge \(l_{s,s'}\) is defined as the closed-path integral of the Berry connection \(\nabla_{\bm{p}}\phi_{s,s'}(\bm{p})\) around the singularity:
	\begin{equation}\label{eq:4}
		l_{s,s'} = \frac{1}{2\pi} \oint_{\mathcal{C}} \nabla_{\bm{p}} \phi_{s,s'}(\bm{p}) \, \mathrm{d}\bm{p} \in \mathbb{Z},
	\end{equation}
	which represents the winding number of the phase around the singularity and is quantized by the single-valuedness of the probability amplitude.

	\section{Numerical results}\label{Sec:III}

	This section presents numerical results for the spin-resolved momentum and phase distributions in multiphoton electron-positron pair production. Throughout Secs.~\ref{Sec:IIIA} and~\ref{SecIII:B}, we set the longitudinal momentum \(p_z = 0\). In this case, the diagonal spinor inner-product terms for the cross-spin configurations vanish, i.e., \(\Delta_{+,-'}(\bm{p},t) = \Delta_{-,+'}(\bm{p},t) = 0\), so that the evolution equations for \(c_{+,-'}^{\beta}(\bm{p},t)\) and \(c_{-,+'}^{\beta}(\bm{p},t)\) reduce to homogeneous form. With the zero initial conditions \(c_{+,-'}^{\beta}(\bm{p},-\infty) = c_{-,+'}^{\beta}(\bm{p},-\infty) = 0\), these homogeneous equations yield \(c_{+,-'}^{\beta}(\bm{p},t) = c_{-,+'}^{\beta}(\bm{p},t) = 0\), and hence \(f_{+,-'}(\bm{p}) = f_{-,+'}(\bm{p}) = 0\). Therefore, for \(p_z = 0\), only the spin configurations \((+,+')\) and \((-,-')\) contribute to the momentum distribution. For \(p_z \neq 0\), however, these inner-product terms become nonzero for all four spin configurations, which will be explored in the analysis of transverse asymmetry in Sec.~\ref{SecIII:C}.

	\subsection{Single elliptically polarized pulse}\label{Sec:IIIA}

\begin{figure}[htbp]
		\centering
		\setlength{\tabcolsep}{0pt}
		\begin{tabular}{@{}ccc@{}}
			\includegraphics[width=0.33\textwidth]{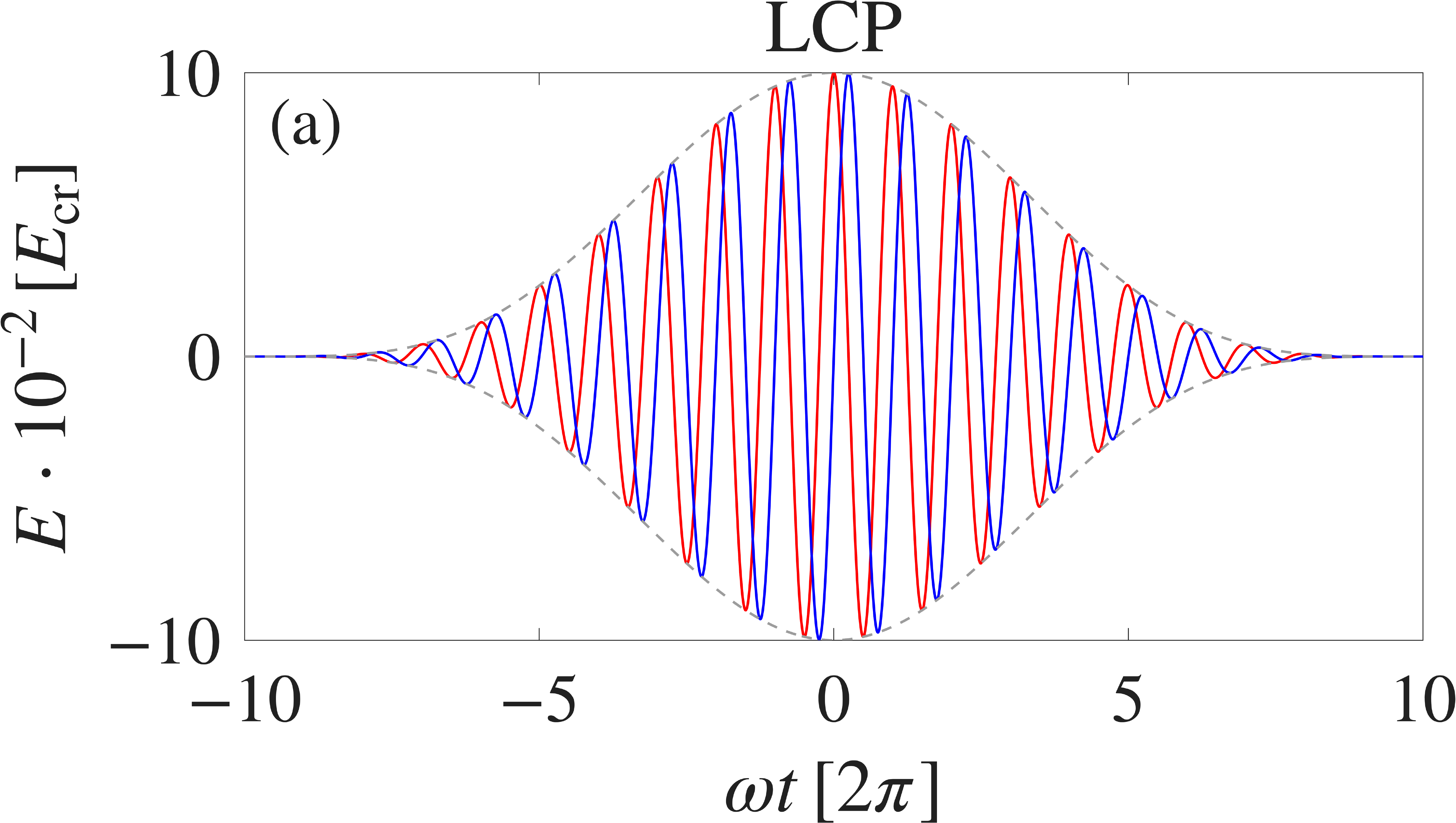} &
			\includegraphics[width=0.33\textwidth]{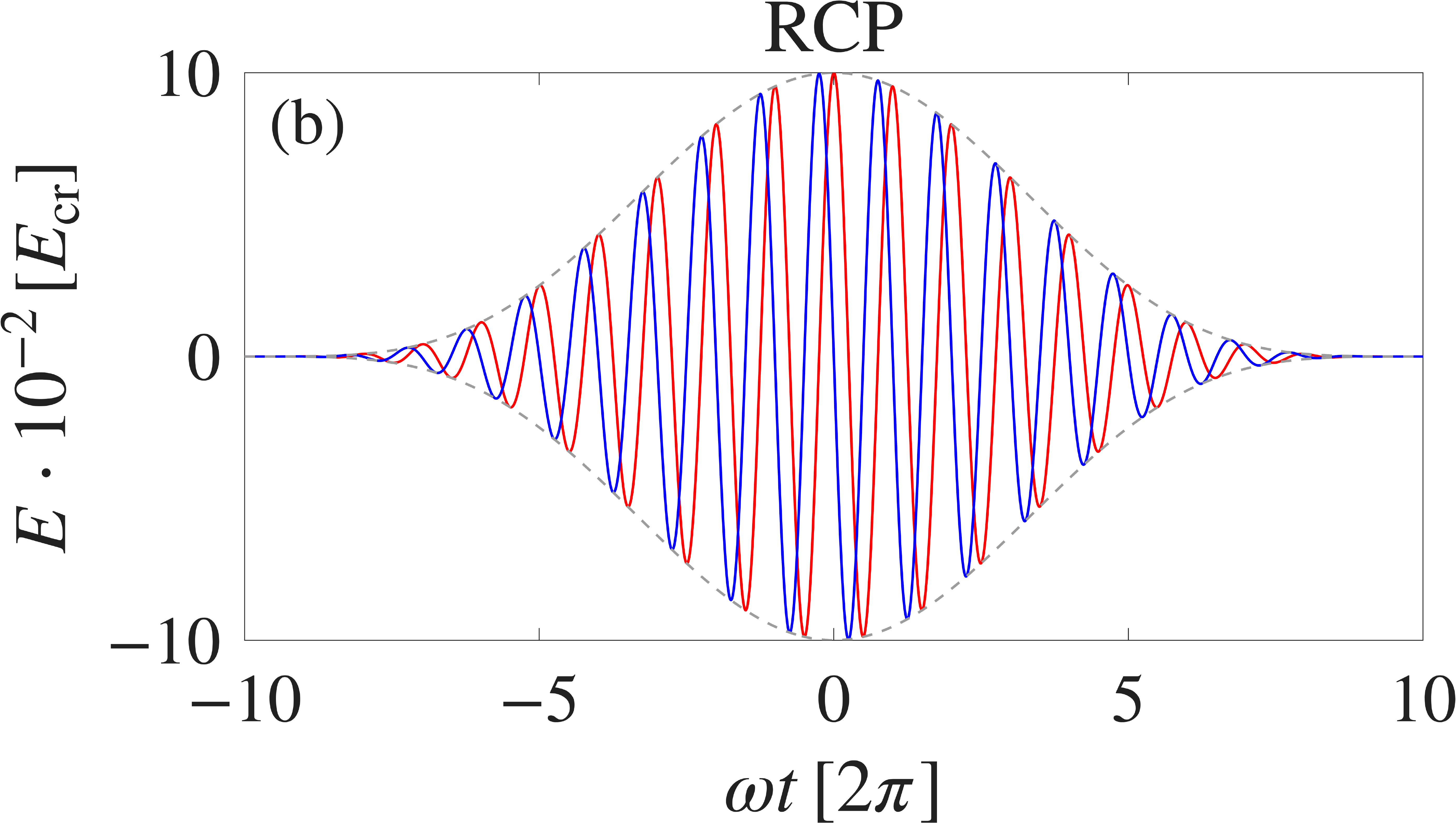} &
			\includegraphics[width=0.33\textwidth]{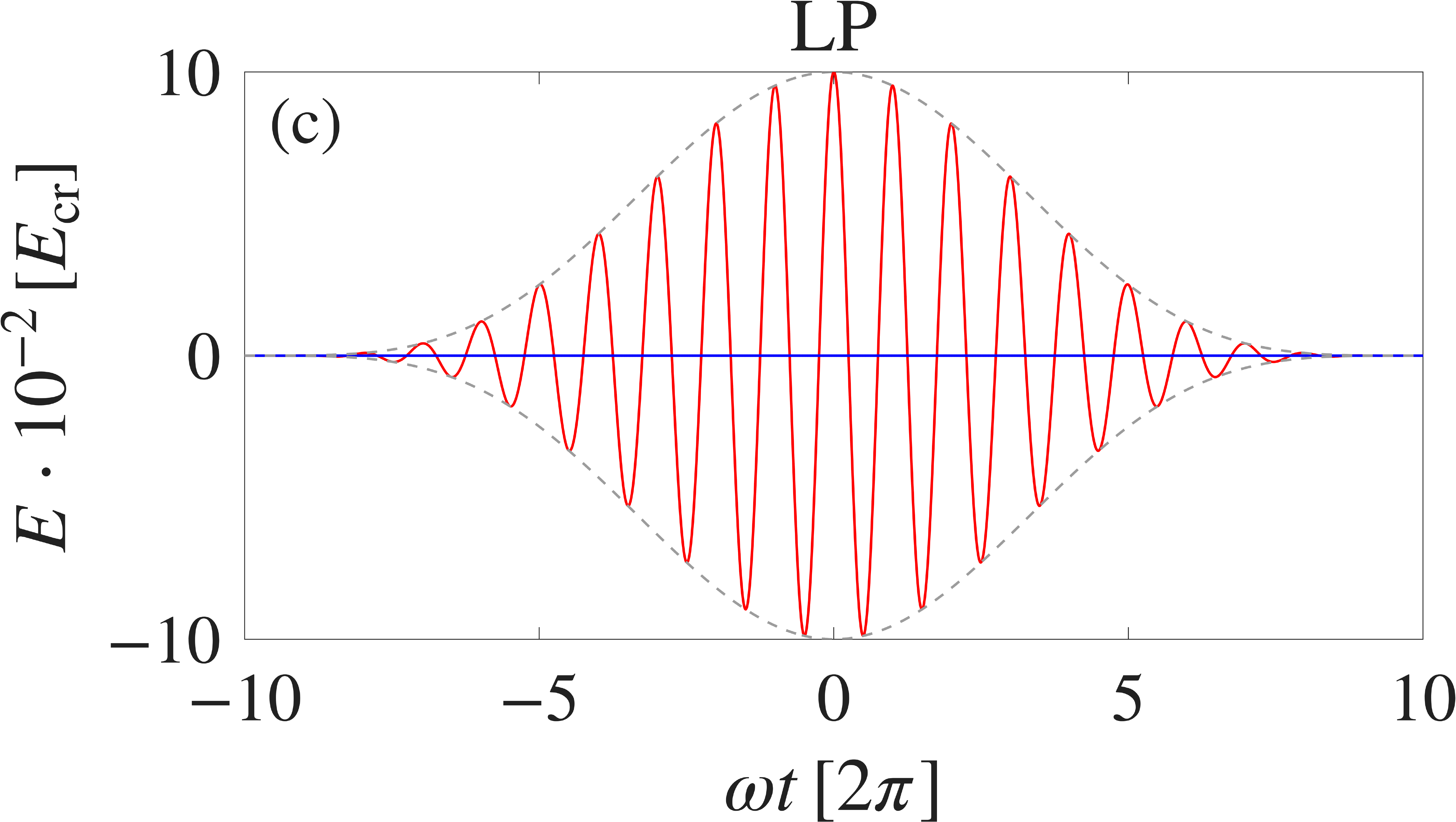}
		\end{tabular}
		\caption{Electric fields for LCP (a), RCP (b), and LP (c). The red and green solid curves denote the \(E_x\) and \(E_y\) components, respectively, and the gray dashed curves denote the envelope. The parameters are \(E_0=0.10E_{\mathrm{cr}}\), \(\tau=20/m\), and \(\omega=0.75m\).}
		\label{Fig:1}
	\end{figure}

\begin{figure}[htbp]
		\centering
		\setlength{\tabcolsep}{0pt}
		\begin{tabular}{@{}c@{}}
			\includegraphics[width=\textwidth]{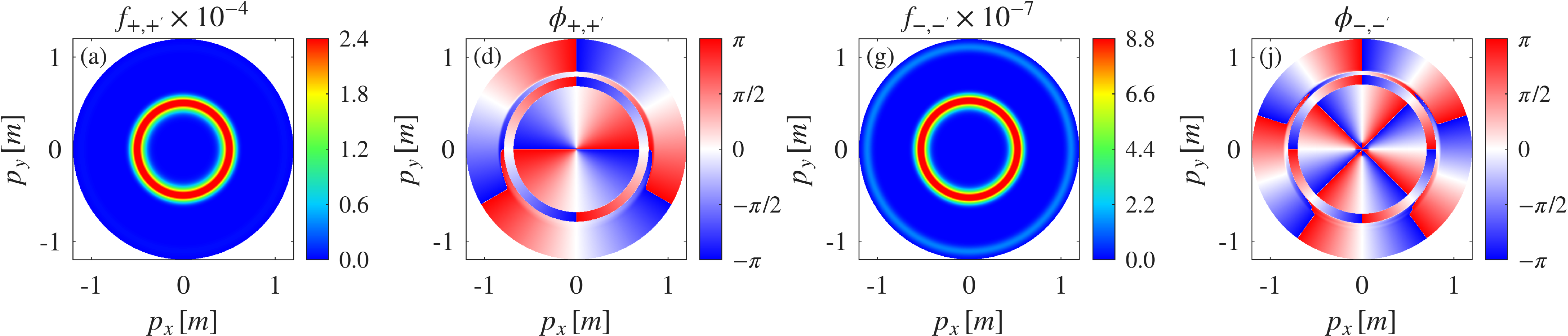} \\[1ex]
			\includegraphics[width=\textwidth]{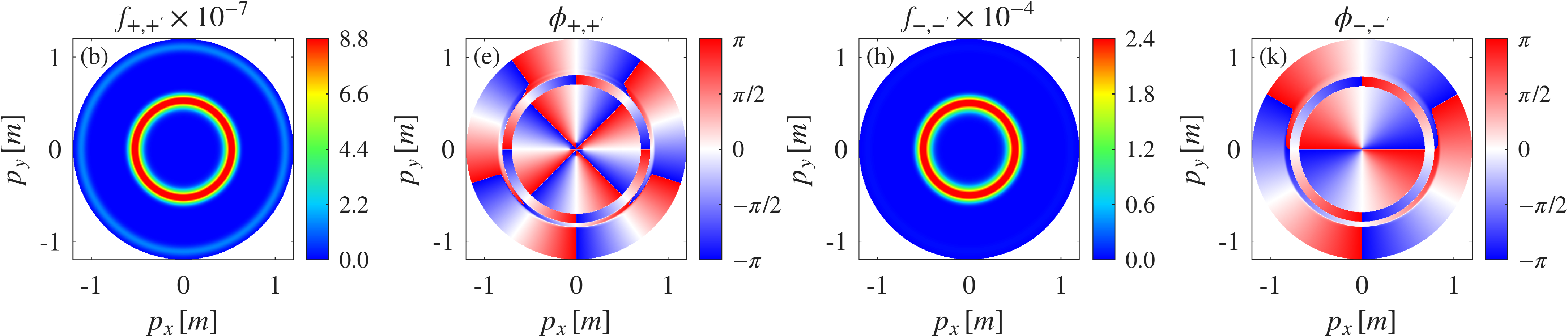} \\[1ex]
			\includegraphics[width=\textwidth]{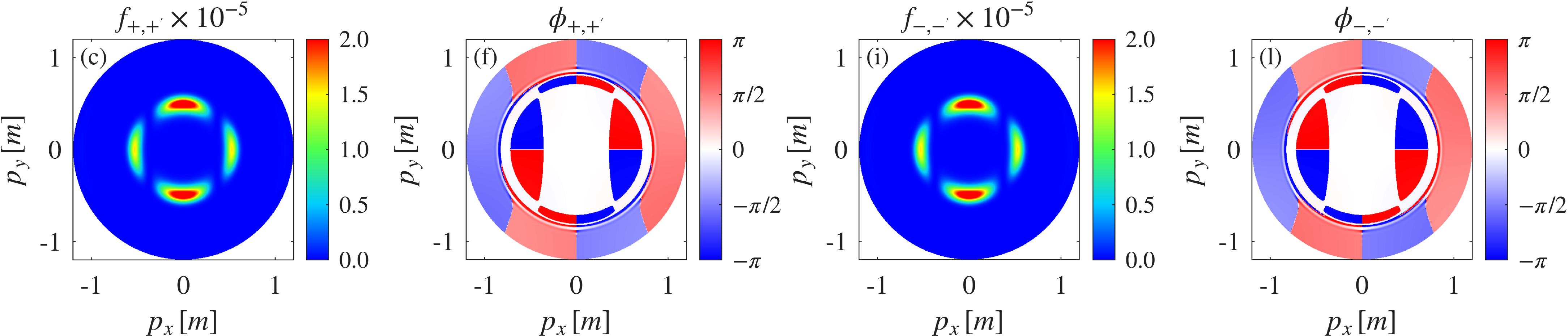}
		\end{tabular}
		\caption{Spin-resolved momentum distributions \(f_{s,s'}\) and phase distributions \(\phi_{s,s'}\) for the spin configurations \((+,+')\) (columns 1 and 2) and \((-,-')\) (columns 3 and 4). The top, middle, and bottom rows correspond to ellipticity \(\delta=+1\), \(-1\), and \(0\), respectively. The parameters are \(E_0=0.10E_{\mathrm{cr}}\), \(\tau=20/m\), \(\omega=0.75m\), and \(p_z=0\).}
		\label{Fig:2}
	\end{figure}

	We first consider a single elliptically polarized electric field
	\begin{equation}\label{eq:5}
		\bm{E}(t) = E_{0}\cos^{4}\left(\frac{\omega t}{2\tau}\right)\left[\cos \left(\omega t\right)\mathbf{e}_{x} + \delta \sin \left(\omega t\right)\mathbf{e}_{y}\right],
	\end{equation}
	where \(\tau\) is the pulse duration and \(\delta\) is the ellipticity. The cases \(\delta = +1, -1\) and \(0\) correspond to left-circular polarization (LCP), right-circular polarization (RCP), and linear polarization (LP), respectively. We set \(E_{0} = 0.10E_{\mathrm{cr}}\), \(\tau = 20/m\), \(\omega = 0.75m\), and \(p_{z} = 0\), where \(E_{\mathrm{cr}} = m^{2} / |e| \approx 1.32\times 10^{18}\mathrm{V/m}\) is the Schwinger critical field strength. The Keldysh adiabaticity parameter \(\gamma_{k} \approx 7.5 \gg 1\) indicates the multiphoton-dominated regime. Figure~\ref{Fig:1} shows the electric fields for LCP, RCP, and LP, and Figure~\ref{Fig:2} shows the corresponding spin-resolved momentum distributions \(f_{s,s'}(\bm{p})\) and phase distributions \(\phi_{s,s'}(\bm{p})\). For circularly polarized fields, the momentum spectra exhibit concentric multiphoton rings centered at the origin, and the phase distributions exhibit quantized vortices, whose cores coincide with the phase singularities. In particular, the momentum-space origin is a phase singularity.

	The quantized vortices are a direct topological signature of the angular momentum transfer from the field to the pairs. The topological charge represents the \(z\)-component of the IOAM of the electron-positron pair. The \(z\)-component of the total angular momentum is conserved in pair production. In a semiclassical picture, each absorbed photon carries \(\pm 1\) unit of angular momentum along the \(z\)-axis, so that an \(N\)-photon absorption process transfers a total angular momentum of \(\pm N\). The transferred angular momentum is composed of the spin projection \(S_{s,s'}\) and the orbital component \(l_{s,s'}\). Consequently, the topological charge for circularly polarized fields satisfies
	\begin{equation}\label{eq:6}
		l_{s,s'} = \mathrm{sgn}(\delta)N - S_{s,s'},
	\end{equation}
	where \(N = \left\lceil \frac{2m_{*}}{\omega} \right\rceil\) is the minimum number of absorbed photons, and \(m_{*} = m\sqrt{1 + \gamma_{k}^{-2}}\) is the effective mass. The sign of \(l_{s,s'}\) distinguishes vortices from antivortices, with opposite winding for LCP and RCP. Specifically, LCP yields clockwise quantized vortices with \(l_{+,+'} = 2\) and \(l_{-,-'} = 4\), whereas RCP yields counterclockwise quantized vortices with \(l_{+,+'} = -4\) and \(l_{-,-'} = -2\).
	
	Under handedness reversal, the spin configurations for LCP and RCP are interchanged, as required by \(C\)-parity conservation: \(f_{+,+'}^{\mathrm{LCP}} = f_{-,-'}^{\mathrm{RCP}}\) and \(f_{-,-'}^{\mathrm{LCP}} = f_{+,+'}^{\mathrm{RCP}}\). For LCP, the \((+,+')\) configuration exhibits a larger momentum distribution but a smaller topological charge magnitude than the \((-,-')\) configuration: \(f_{+,+'}^{\mathrm{LCP}} > f_{-,-'}^{\mathrm{LCP}}\) and \(|l_{+,+'}^{\mathrm{LCP}}| < |l_{-,-'}^{\mathrm{LCP}}|\). For RCP, the inequalities reverse: \(f_{+,+'}^{\mathrm{RCP}} < f_{-,-'}^{\mathrm{RCP}}\) and \(|l_{+,+'}^{\mathrm{RCP}}| > |l_{-,-'}^{\mathrm{RCP}}|\). This inverse correlation favors lower-IOAM pair production, consistent with angular momentum conservation.
	
	The LP field is a superposition of equal-amplitude LCP and RCP components, each with half of its amplitude. Due to \(C\)-parity conservation and the equal-weight superposition of the LCP and RCP components, the \((+,+')\) and \((-,-')\) configurations yield identical momentum distributions in the LP field, i.e., \(f_{+,+'}^{\mathrm{LP}} = f_{-,-'}^{\mathrm{LP}}\). Since the LP field oscillates purely along the \(x\)-axis, the momentum distribution is symmetric under the reflections \(p_x \rightarrow - p_x\) and \(p_y \rightarrow - p_y\). The interference between the LCP and RCP multiphoton pathways yields an azimuthally dependent relative phase, which modulates the multiphoton rings into a four-lobe structure. Unlike the quantized vortices in circularly polarized fields, the LP phase distribution exhibits nodal lines, with phase jumps of \(\pi\) across each line. The closed-path integral around a nodal line vanishes, which corresponds to zero topological charge. Since the LCP and RCP angular momenta mutually cancel, the net OAM of the electron-positron pair vanishes, with no quantized vortices formed.

	\subsection{Two circularly polarized pulses with a time delay}\label{SecIII:B}

	To investigate vortex evolution, we extend the analysis to two circularly polarized pulses with a controllable relative time delay
	\begin{equation}\label{eq:7}
		\bm{E}(t) = E_0 \cos^4\left(\frac{\omega t}{2\tau}\right)\begin{pmatrix} \cos(\omega t) \\ \delta_1 \sin(\omega t) \\ 0 \end{pmatrix} + E_0 \cos^4\left[\frac{\omega(t - T)}{2\tau}\right]\begin{pmatrix} \cos[\omega(t - T)] \\ \delta_2 \sin[\omega(t - T)] \\ 0 \end{pmatrix},
	\end{equation}
	where \(\delta_{1,2} = \pm 1\) are the circular polarization parameters, and \(T \in [0,4T_p]\) is the relative time delay, with \(T_p = \pi \tau / \omega\) the pulse half-width. We consider the left-left circular polarization (LLCP) configuration with \(\delta_1 = \delta_2 = 1\) and the right-left circular polarization (RLCP) configuration with \(\delta_1 = -\delta_2 = -1\). The parameters are \(E_0 = 0.05E_{\mathrm{cr}}\), \(\tau = 20/m\), \(\omega = 0.75m\), and \(p_z = 0\), corresponding to a Keldysh adiabaticity parameter \(\gamma_k \approx 15 \gg 1\), indicating the multiphoton-dominated regime.
	
	For the two-pulse configuration, the probability amplitude is a coherent superposition of the single-pulse amplitudes:
	\begin{equation}\label{eq:8}
		c_{s,s'}^{\beta}(\bm{p}) = c_{s,s'}^{\beta,(1)}(\bm{p}) + c_{s,s'}^{\beta,(2)}(\bm{p})\mathrm{e}^{\mathrm{i}\omega_{\bm{p}}T} + \mathcal{C}_{s,s'}(\bm{p}),
	\end{equation}
	where \(c_{s,s'}^{\beta,(1)}(\bm{p})\) and \(c_{s,s'}^{\beta,(2)}(\bm{p})\) are the single-pulse probability amplitudes, \(\mathcal{C}_{s,s'}(\bm{p})\) is the inter-pulse correlation amplitude, and \(\omega_{\bm{p}}T\) is the relative phase induced by the time delay, whose rigorous derivation is provided in Appendix~\ref{App:B}. The corresponding spin-resolved momentum distribution is
	\begin{equation}\label{eq:9}
		\begin{aligned}
			f_{s,s'}(\bm{p}) &= f_{s,s'}^{(1)}(\bm{p}) + f_{s,s'}^{(2)}(\bm{p}) \\
			&\quad + 2\mathrm{Re} \left[ c_{s,s'}^{\beta,(1)*}(\bm{p}) c_{s,s'}^{\beta,(2)}(\bm{p}) \mathrm{e}^{\mathrm{i}\omega_{\bm{p}}T} \right] \\
			&\quad + 2\mathrm{Re} \left[ \left( c_{s,s'}^{\beta,(1)*}(\bm{p}) + c_{s,s'}^{\beta,(2)*}(\bm{p}) \mathrm{e}^{-\mathrm{i}\omega_{\bm{p}}T} \right) \mathcal{C}_{s,s'}(\bm{p}) \right] \\
			&\quad + |\mathcal{C}_{s,s'}(\bm{p})|^2 ,
		\end{aligned}
	\end{equation}
	where \(f_{s,s'}^{(1)}(\bm{p})\) and \(f_{s,s'}^{(2)}(\bm{p})\) denote the momentum distributions for the two pulses acting alone. The remaining terms are the coherent, cross, and correlation terms: the coherent term represents the interference between the two single-pulse amplitudes, the cross term arises from interference between the correlation amplitude and the single-pulse amplitudes, and the correlation term is the modulus squared of the correlation amplitude.
	
\begin{figure}[htbp]
		\centering
		\setlength{\tabcolsep}{0pt}
		\begin{tabular}{@{}ccc@{}}
			\includegraphics[width=0.33\textwidth]{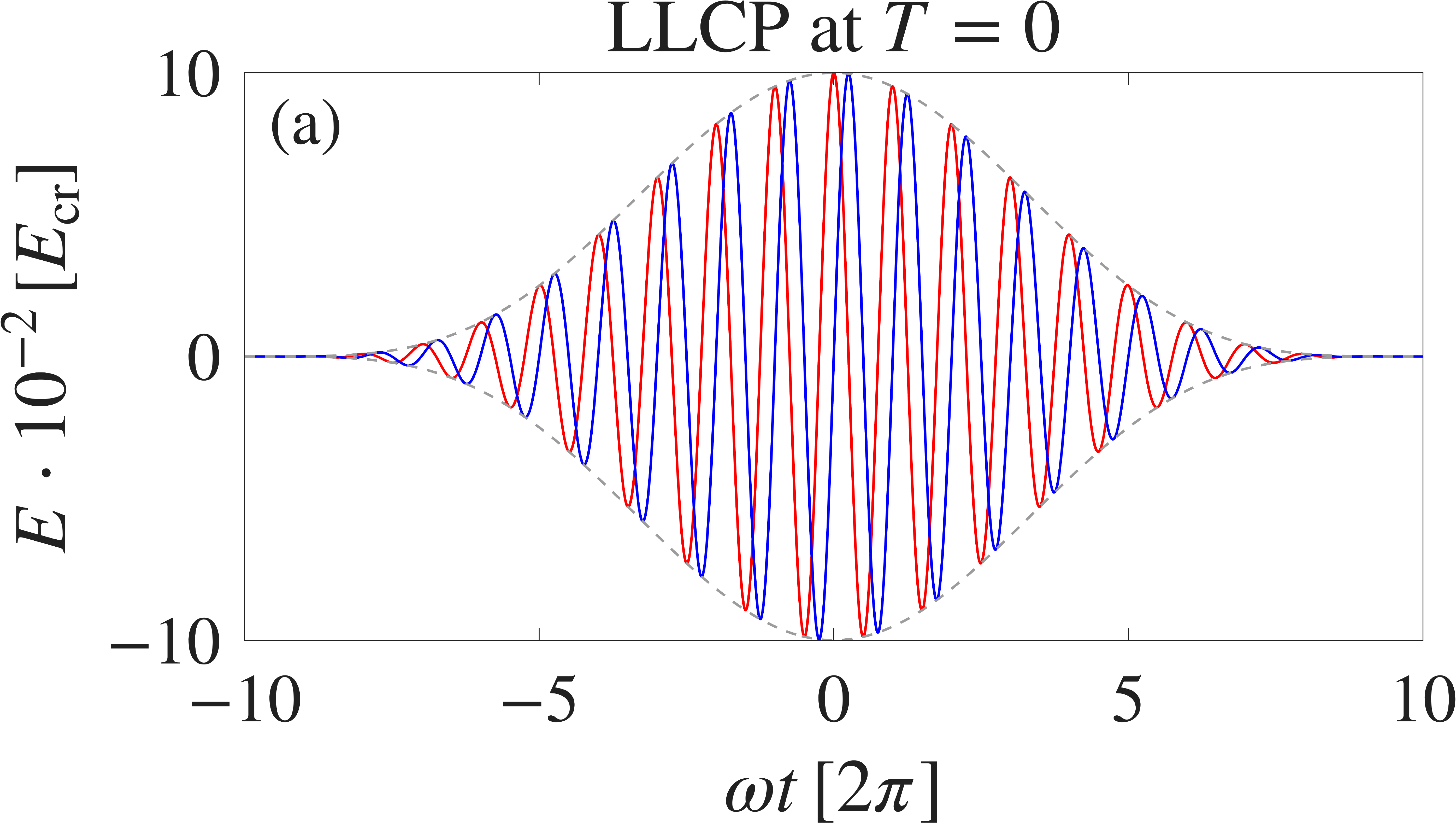} &
			\includegraphics[width=0.33\textwidth]{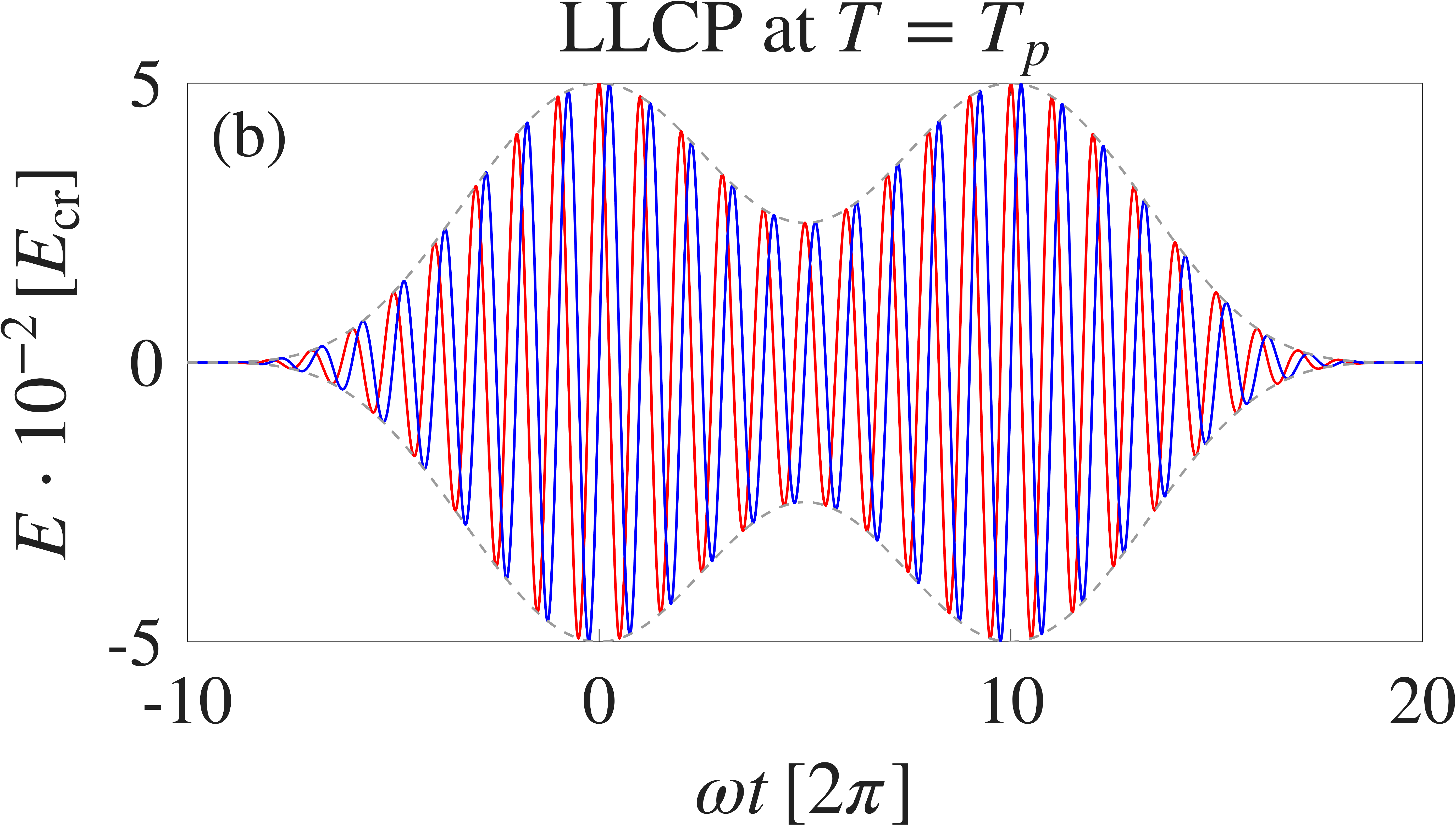} &
			\includegraphics[width=0.33\textwidth]{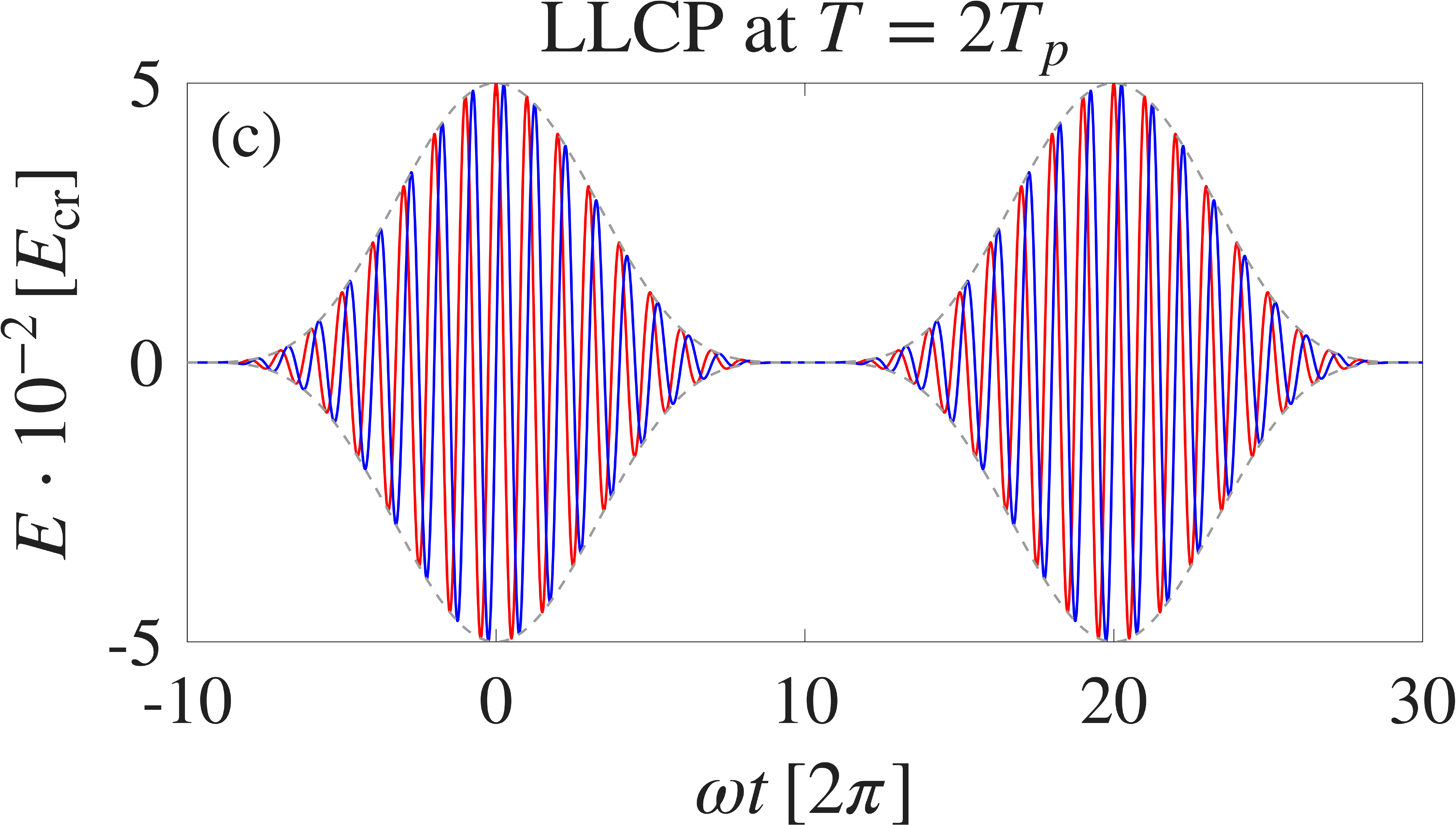}
		\end{tabular}
		\caption{Electric fields for LLCP at time delays of \(T=0\) (a), \(T_p\) (b), and \(2T_p\) (c). The red and green solid curves denote the \(E_x\) and \(E_y\) components, respectively, and the gray dashed curves denote the envelope. The parameters are \(E_0=0.05E_{\mathrm{cr}}\), \(\tau=20/m\), and \(\omega=0.75m\).}
		\label{Fig:3}
	\end{figure}

\begin{figure}[htbp]
		\centering
		\setlength{\tabcolsep}{0pt}
		\begin{tabular}{@{}c@{}}
			\includegraphics[width=\textwidth]{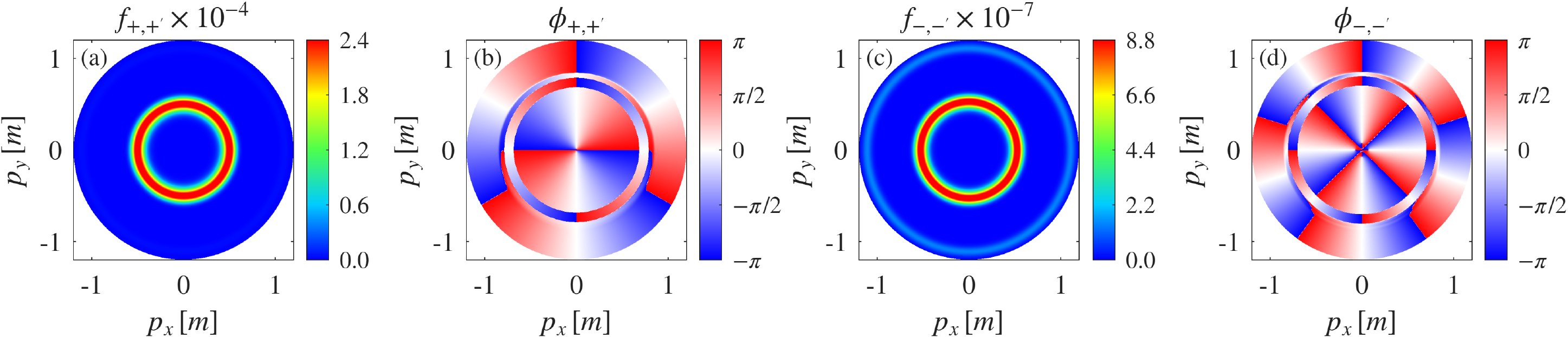} \\[1ex]
			\includegraphics[width=\textwidth]{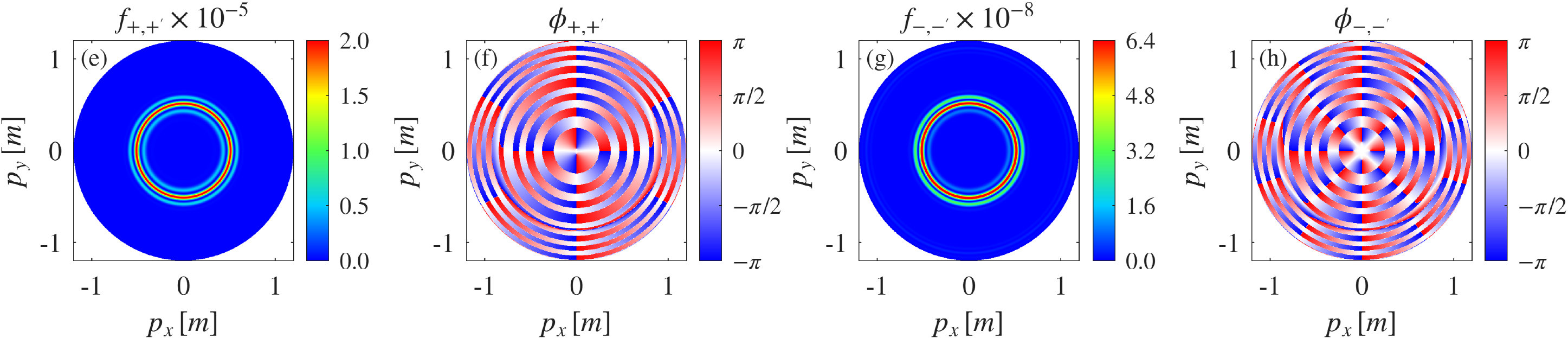} \\[1ex]
			\includegraphics[width=\textwidth]{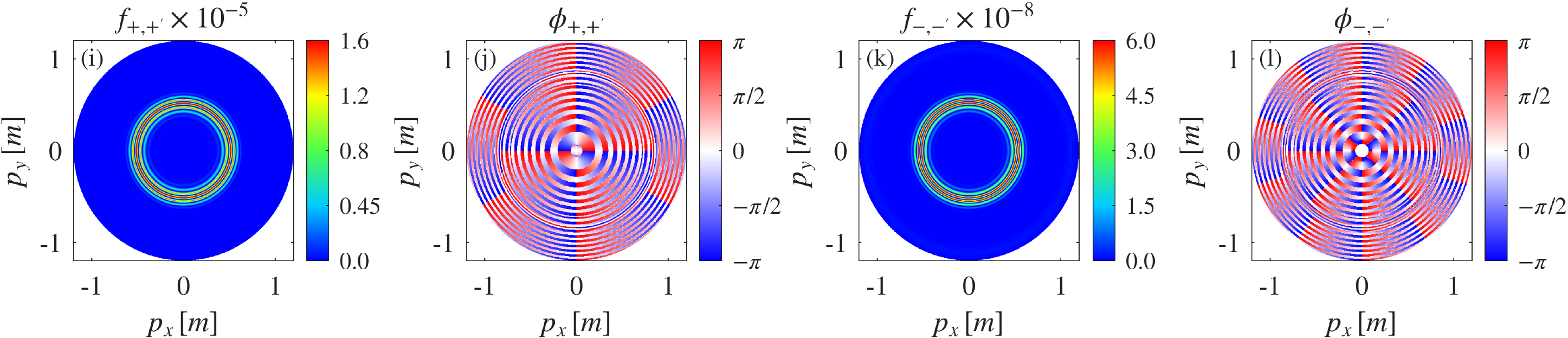}
		\end{tabular}
		\caption{Spin-resolved momentum distributions \(f_{s,s'}\) and phase distributions \(\phi_{s,s'}\) for LLCP. Columns 1 and 2 correspond to the spin configuration \((+,+')\), and columns 3 and 4 correspond to \((-,-')\). The top, middle, and bottom rows correspond to time delays \(T=0\), \(T_p\), and \(2T_p\), respectively. The parameters are \(E_0=0.05E_{\mathrm{cr}}\), \(\tau=20/m\), \(\omega=0.75m\), and \(p_z=0\).}
		\label{Fig:4}
	\end{figure}

	Figure~\ref{Fig:3} shows the LLCP field at time delays \(T = 0\), \(T_p\), and \(2T_p\). Figure~\ref{Fig:4} shows the corresponding spin-resolved momentum distributions \(f_{s,s'}(\bm{p})\) and phase distributions \(\phi_{s,s'}(\bm{p})\). The LLCP field is a superposition of two co-rotating LCP pulses separated by a time delay \(T\). Across all time delays, the \((+,+')\) configuration always dominates over the \((-,-')\) configuration in the momentum distribution.

	At \(T = 0\), the two pulses fully overlap and add in phase, forming a single LCP pulse of amplitude \(2E_0\). Thus, LLCP with \(E_0 = 0.05E_{\mathrm{cr}}\) yields momentum and phase distributions strictly identical to those of a single LCP pulse with \(E_0 = 0.10E_{\mathrm{cr}}\). The \((+,+')\) configuration exhibits a single central ring in the momentum spectrum, whereas the \((-,-')\) configuration exhibits an additional weak outer ring.
	
	For \(T > 0\), the time delay introduces a relative phase that induces quantum interference, thereby driving the evolution of the momentum and phase distributions. The inter-pulse correlation term \(\mathcal{C}_{s,s'}(\bm{p})\) also contributes to the phase structure, but its effect diminishes as the pulses separate. As \(T\) increases, the momentum peak decreases, the central ring narrows, and new interference rings appear successively inside and outside it, evolving the momentum spectrum from a single ring into dense multi-ring Ramsey fringes, while the number of vortex rings in the phase distribution increases correspondingly. This trend follows from \(\partial (\omega_{\bm{p}}T) / \partial \bm{p} = \bm{p}T / \omega_{\bm{p}}\), which implies that increasing \(T\) makes the phase vary faster with \(\bm{p}\), thereby enhancing the density of phase singularities in momentum space. For \(T \geq T_p\), the two co-rotating LCP pulses have identical momentum distributions in each spin configuration, and their coherent addition doubles the phase winding at the central singularity. Consequently, the topological charges of the LLCP field double those of the individual pulses, namely, \(l_{+,+'} = 4\) and \(l_{-,-'} = 8\).
	
\begin{figure}[htbp]
		\centering
		\setlength{\tabcolsep}{0pt}
		\begin{tabular}{@{}ccc@{}}
			\includegraphics[width=0.33\textwidth]{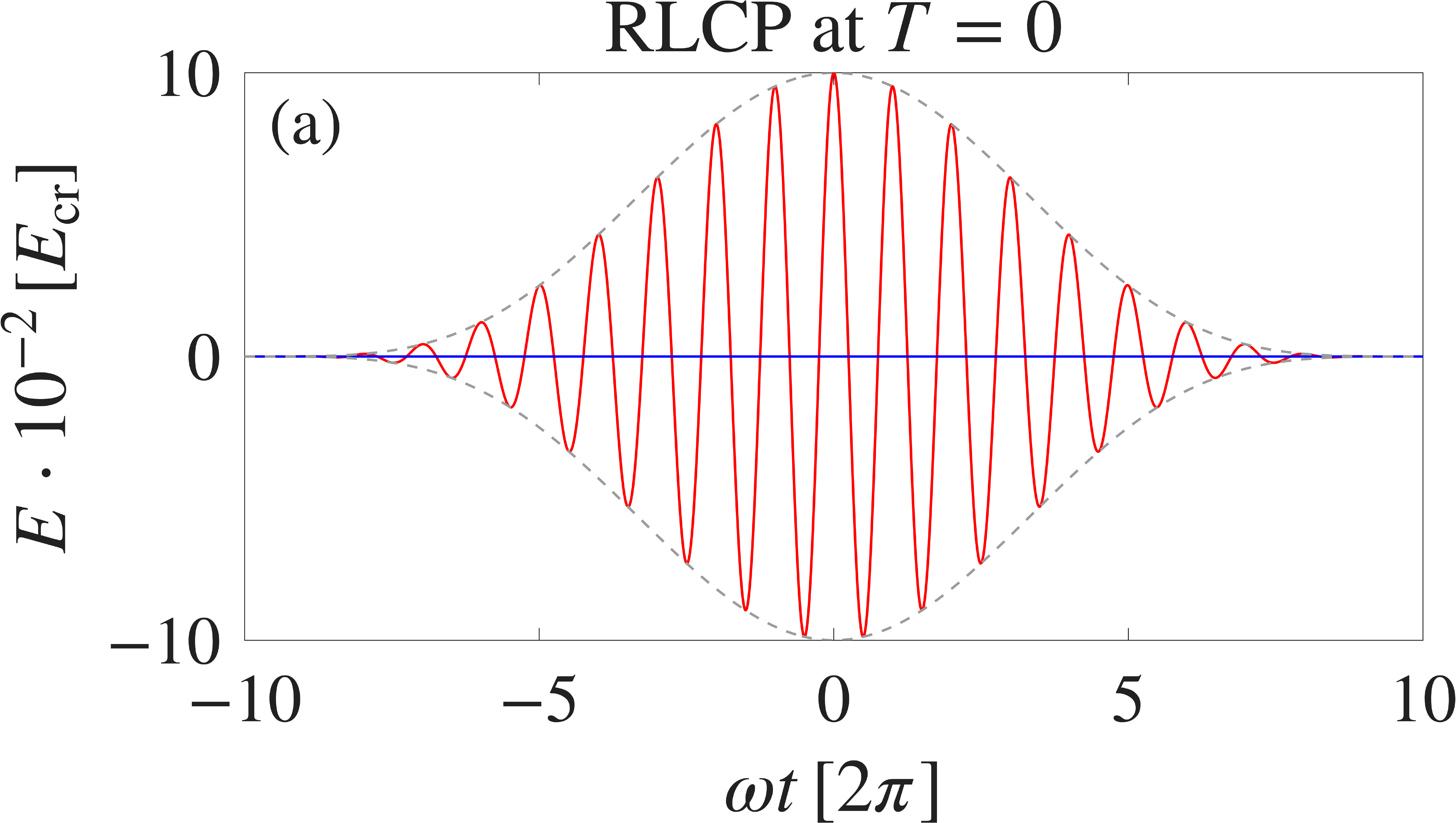} &
			\includegraphics[width=0.33\textwidth]{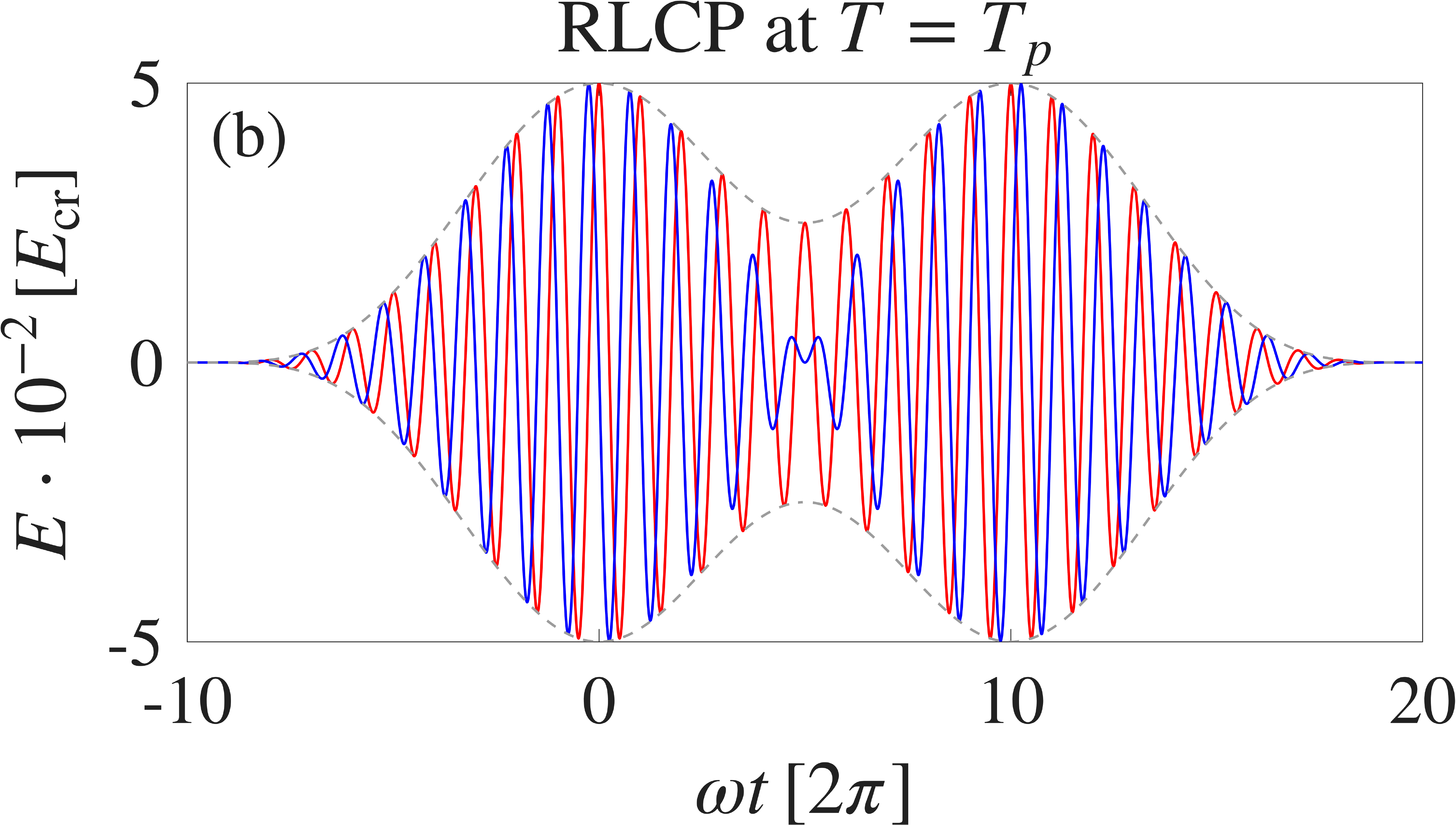} &
			\includegraphics[width=0.33\textwidth]{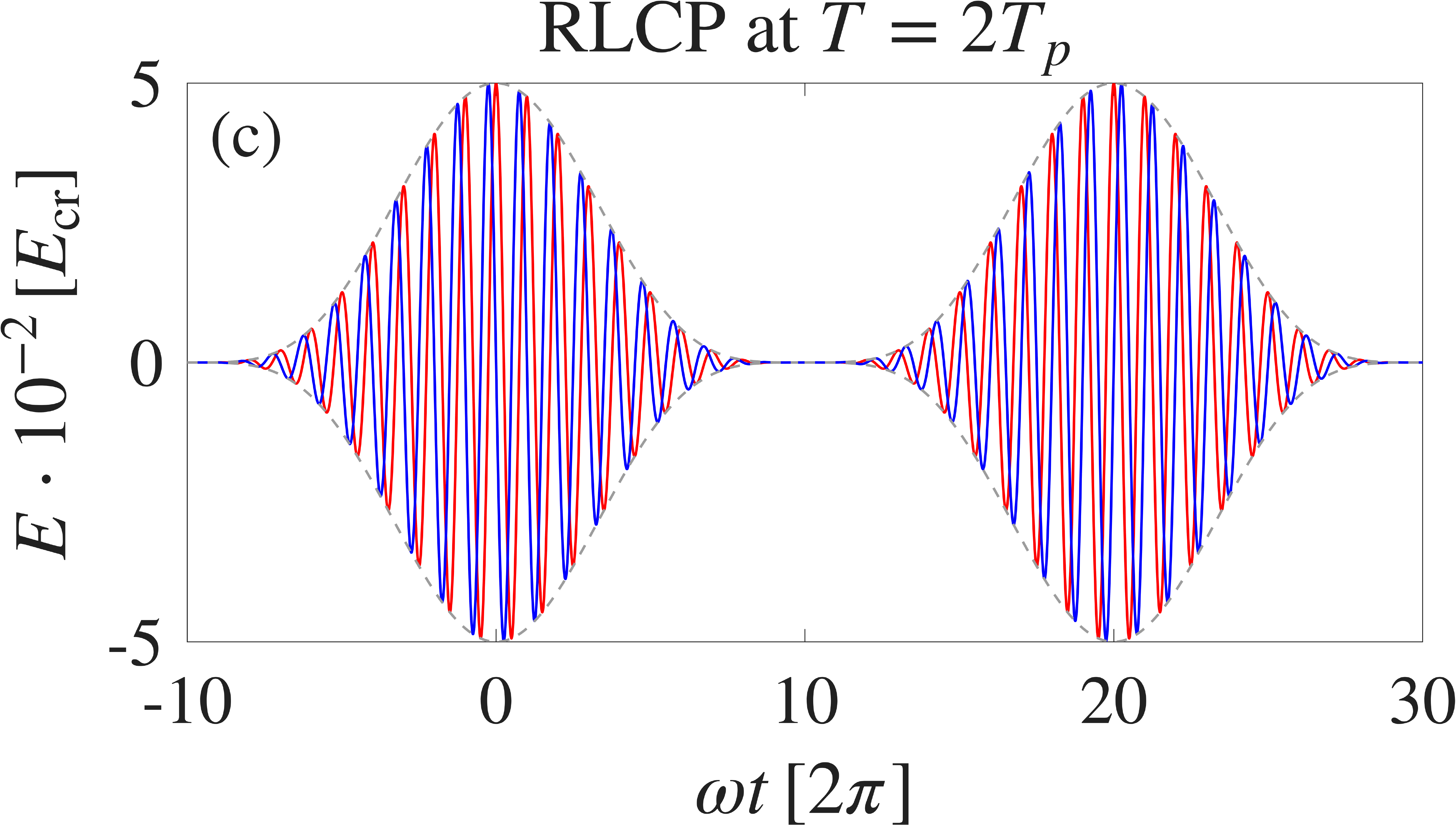}
		\end{tabular}
		\caption{Electric fields for RLCP at time delays of \(T=0\) (a), \(T_p\) (b), and \(2T_p\) (c). The red and green solid curves denote the \(E_x\) and \(E_y\) components, respectively, and the gray dashed curves denote the envelope. The parameters are \(E_0=0.05E_{\mathrm{cr}}\), \(\tau=20/m\), and \(\omega=0.75m\).}
		\label{Fig:5}
	\end{figure}

\begin{figure}[htbp]
		\centering
		\setlength{\tabcolsep}{0pt}
		\begin{tabular}{@{}c@{}}
			\includegraphics[width=\textwidth]{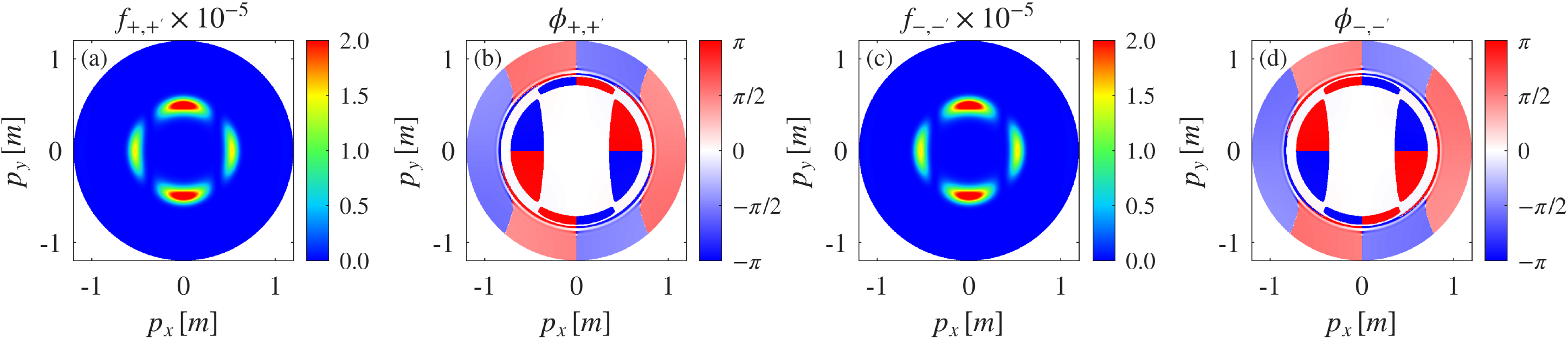} \\[1ex]
			\includegraphics[width=\textwidth]{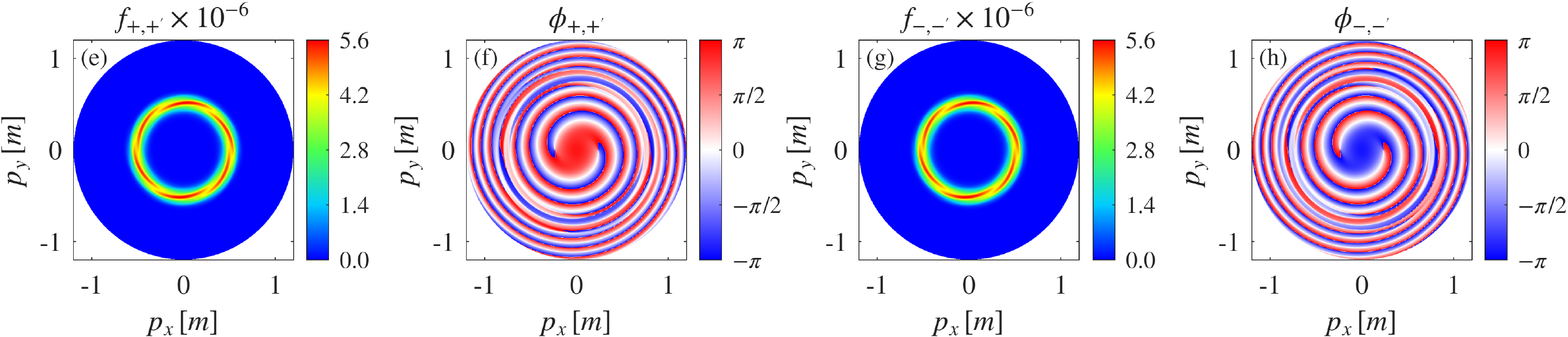} \\[1ex]
			\includegraphics[width=\textwidth]{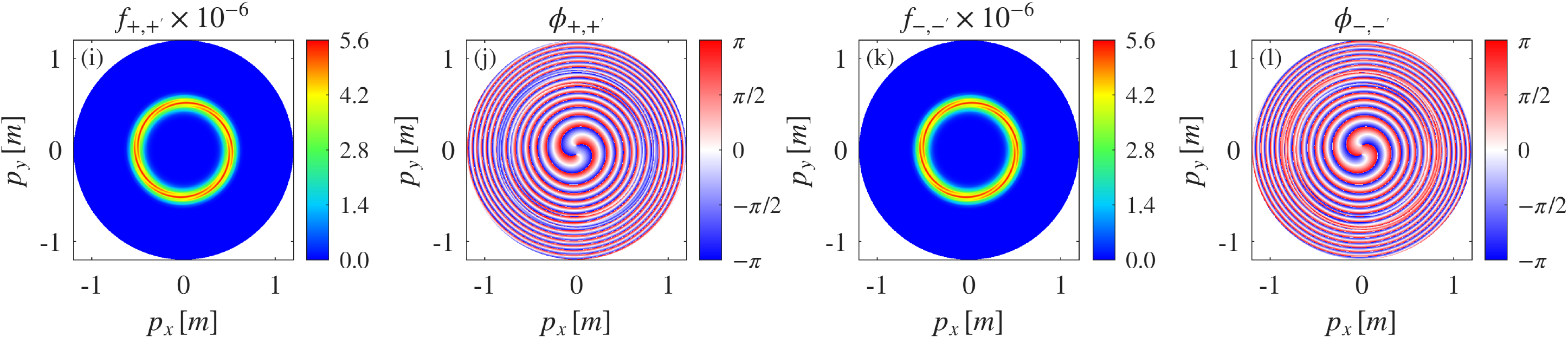}
		\end{tabular}
		\caption{Spin-resolved momentum distributions \(f_{s,s'}\) and phase distributions \(\phi_{s,s'}\) for RLCP. Columns 1 and 2 correspond to the spin configuration \((+,+')\), and columns 3 and 4 correspond to \((-,-')\). The top, middle, and bottom rows correspond to time delays \(T=0\), \(T_p\), and \(2T_p\), respectively. The parameters are \(E_0=0.05E_{\mathrm{cr}}\), \(\tau=20/m\), \(\omega=0.75m\), and \(p_z=0\).}
		\label{Fig:6}
	\end{figure}

	Figure~\ref{Fig:5} shows the RLCP field at time delays \(T = 0\), \(T_p\), and \(2T_p\). Figure~\ref{Fig:6} shows the corresponding spin-resolved momentum distributions \(f_{s,s'}(\bm{p})\) and phase distributions \(\phi_{s,s'}(\bm{p})\). The RLCP field is a coherent superposition of two counter-rotating circularly polarized pulses, one RCP and one LCP, separated by a time delay \(T\). Across all time delays, the \((+,+')\) and \((-,-')\) configurations have identical momentum distributions, as required by \(C\)-parity conservation and the equal-weight superposition of the two components.

	At \(T = 0\), the two pulses fully overlap. Since \(\delta_1 = -\delta_2\), their \(E_y\) components cancel, leaving a linearly polarized pulse of amplitude \(2E_0\). Therefore, RLCP with \(E_0 = 0.05E_{\mathrm{cr}}\) is identical to a single LP pulse with \(E_0 = 0.10E_{\mathrm{cr}}\) in both momentum and phase distributions. As a linearly polarized field exhibits nodal lines rather than quantized vortices, the topological charges vanish, namely, \(l_{+,+'} = 0\) and \(l_{-,-'} = 0\).
	
	When \(T \geq T_p\), each pulse effectively contributes three absorbed photons, giving a total of six absorbed photons. The momentum spectrum therefore exhibits a six-arm spiral vortex structure, with the number of arms equal to the total number of absorbed photons. Since the RCP and LCP components contribute differently to each spin configuration, the dominant component in the momentum distribution determines the topological charges of the RLCP field. In the \((+,+')\) configuration, the LCP component dominates over the RCP component in the momentum distribution, hence this configuration carries the topological charge \(l_{+,+'} = 2\), with the winding direction following the LCP rotation. Conversely, in the \((-,-')\) configuration, the RCP component dominates, giving \(l_{-,-'} = -2\), with the winding direction following the RCP rotation.
	
	To quantify the interference between the two pulses induced by the time delay, we define the interference degree as
	\begin{equation}\label{eq:10}
		I_{s,s'} = \frac{f_{s,s'} - f_{s,s'}^{(1)} - f_{s,s'}^{(2)}}{f_{s,s'}}.
	\end{equation}
	\(I_{s,s'} > 0\), \(I_{s,s'} < 0\), and \(I_{s,s'} = 0\) correspond to constructive interference, destructive interference, and the absence of interference, respectively. The numerator is the sum of the coherent, cross, and correlation terms defined in Eq.~\eqref{eq:9}, whose interplay governs the evolution of \(I_{s,s'}\).
	
\begin{figure}[htbp]
		\centering
		\includegraphics[scale=0.6]{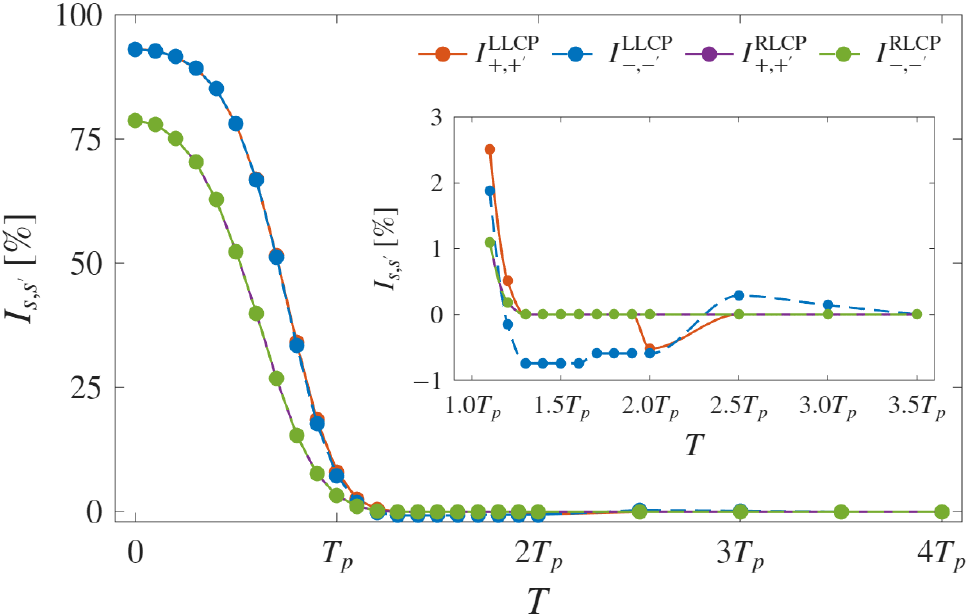}
		\caption{Interference degree \(I_{s,s'}\) versus the time delay \(T\). The red solid and blue dashed curves represent the \((+,+')\) and \((-,-')\) spin configurations of LLCP, respectively, while the purple solid and green dashed curves represent the corresponding spin configurations of RLCP. The inset shows a magnified view for \(T\in[1.1T_p,3.5T_p]\). The parameters are \(E_0=0.05E_{\mathrm{cr}}\), \(\tau=20/m\), \(\omega=0.75m\), and \(p_z=0\).}
		\label{Fig:7}
	\end{figure}

	Figure~\ref{Fig:7} shows the time-delay dependence of the interference degree for LLCP and RLCP. At \(T = 0\), the two pulses fully overlap with vanishing relative phase, thus both the coherent addition of the two single-pulse amplitudes and the inter-pulse correlation reach their maxima, resulting in the most pronounced deviation from the single-pulse sum.
	
	As \(T\) increases, the relative phase \(\omega_{\bm{p}}T\) varies rapidly with momentum. Consequently, the coherent term and the oscillatory part of the cross term average to zero in momentum space, causing the interference degree to decay. The residual interference is subsequently determined by the non-oscillatory part of the cross term and the correlation term. For RLCP, the counter-rotating pulses suppress the nonlinear inter-pulse correlation, rendering \(\mathcal{C}_{s,s'}(\bm{p})\) negligible and leaving no residual interference. Both the \((+,+')\) and \((-,-')\) configurations decay identically to zero for \(T \geq 1.3T_p\). For LLCP, the co-rotating pulses enhance the nonlinear inter-pulse correlation, so that the residual interference is governed by the cross term and the correlation term. Their relative sign and magnitude govern the evolution of the interference for a given spin configuration and time delay. Thus, the chirality combination determines whether residual interference exists, while the spin configuration selects its sign and magnitude, providing a direct control parameter for quantum interference in multiphoton pair production.

	\subsection{Transverse asymmetry}\label{SecIII:C}

	In this section, we analyze the transverse asymmetry of spin-resolved momentum distributions induced by the diagonal spinor inner-product terms. We apply a momentum reflection transformation \(\mathcal{P}: \bm{p} \to \bm{p}^{\mathcal{P}}\) to the mTLM evolution equations. As derived in Appendix~\ref{App:C}, the ratio of the momentum distribution functions is
	\begin{equation}\label{eq:11}
		\frac{f_{s,s'}(\bm{p}^{\mathcal{P}})}{f_{s,s'}(\bm{p})} = \exp \left\{ 2 \int_{-\infty}^{+\infty} \mathrm{Im} \left[ \Delta_{s,s'}(\bm{p}^{\mathcal{P}}, t) - \Delta_{s,s'}(\bm{p}, t) \right] \mathrm{d}t \right\},
	\end{equation}
	which shows that the transverse asymmetry is determined entirely by the imaginary part of the difference between the diagonal spinor inner-product terms.

	\begin{figure}[htbp]
		\centering
		\setlength{\tabcolsep}{0pt}
		\begin{tabular}{@{}c@{}}
			\includegraphics[width=\textwidth]{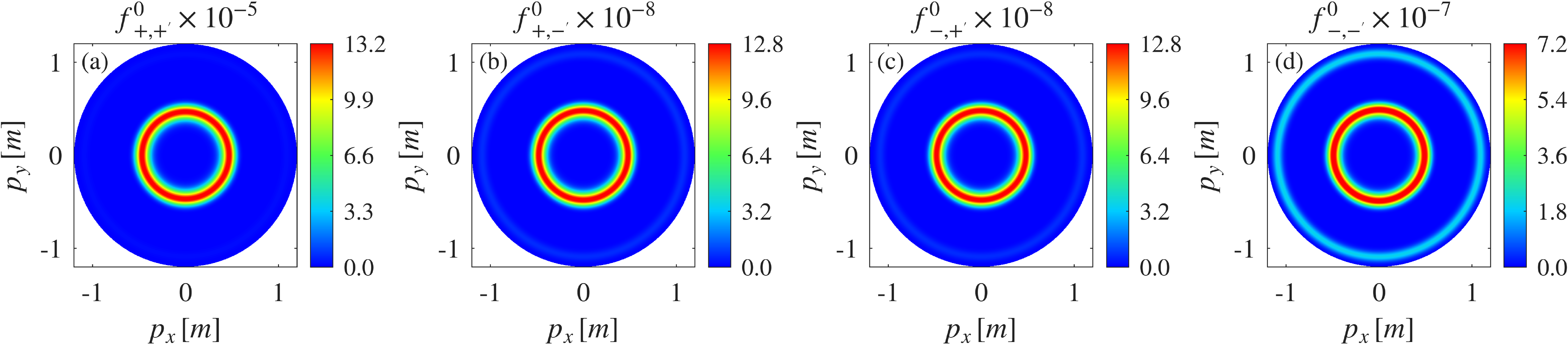} \\[6pt]
			\includegraphics[width=\textwidth]{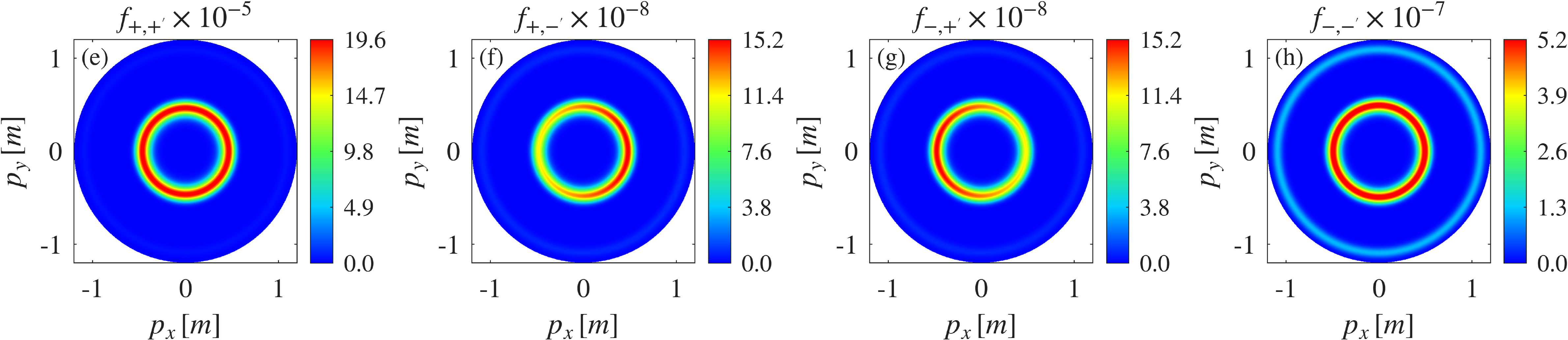}
		\end{tabular}
		\caption{Spin-resolved momentum distributions for the LCP field obtained from the TLM (top row, \(f^0_{s,s'}\)) and the mTLM (bottom row, \(f_{s,s'}\)). Columns 1-4 correspond to the spin configurations \((+,+')\), \((+,-')\), \((-,+')\), and \((-,-')\), respectively. The parameters are \(E_0=0.10E_{\mathrm{cr}}\), \(\tau=20/m\), \(\omega=0.75m\), and \(p_z=0.2\).}
		\label{Fig:8}
	\end{figure}

	Figure~\ref{Fig:8} shows the spin-resolved momentum distributions for the four spin configurations in the LCP field, with the TLM and mTLM distributions denoted by \(f_{s,s'}^{0}(\bm{p})\) and \(f_{s,s'}(\bm{p})\) respectively. In the TLM, with the diagonal spinor inner-product terms neglected, all distributions preserve full-space reflection symmetry, namely, \(f_{s,s'}^{0}(\bm{p}^{\mathcal{P}}) = f_{s,s'}^{0}(\bm{p})\). In the mTLM, the diagonal spinor inner-product terms are retained. Eq.~\eqref{eq:A10} shows that \(\Delta_{+,+'}(\bm{p},t)\) and \(\Delta_{-,-'}(\bm{p},t)\) are purely real, whereas \(\Delta_{+,-'}(\bm{p},t)\) and \(\Delta_{-,+'}(\bm{p},t)\) have nonzero imaginary parts. Consequently, the \((+,+')\) and \((-,-')\) configurations remain symmetric, whereas the \((+,-')\) and \((-,+')\) configurations shift toward positive and negative \(p_x\) respectively. These results demonstrate that the diagonal spinor inner-product terms break the full-space reflection symmetry of the momentum distribution only for the cross-spin configurations.
	
	To quantify the effect of the diagonal spinor inner-product terms on the momentum distribution functions, we define the relative deviation
	\begin{equation}\label{eq:12}
		\kappa_{s,s'} \equiv \frac{f_{s,s'} - f_{s,s'}^{0}}{f_{s,s'}}.
	\end{equation}
	The diagonal spinor inner-product terms significantly modify the momentum distribution functions for each spin configuration: \(\kappa_{+,+'} = +32.35\%\), \(\kappa_{+,-'} = \kappa_{-,+'} = +14.10\%\), and \(\kappa_{-,-'} = -39.32\%\). These deviations originate from the nonadiabatic modulation induced by the diagonal spinor inner-product terms, which are essential for the fine structure of the momentum spectrum. This further establishes the mTLM as a necessary theoretical framework for the accurate description of spin-resolved pair production.

	\section{Summary}\label{Sec:IV}

	We have investigated spin-resolved momentum distributions and phase vortices in multiphoton electron-positron pair production within the mTLM. In particular, we focus on the vortex evolution driven by the time delay between two circularly polarized pulses, while single circularly polarized pulses and single linearly polarized pulses serve as reference cases for vortex formation and annihilation.
	
	We have theoretically derived the spin-resolved probability amplitudes for electron-positron pair production from the Dirac equation. The mTLM retains the diagonal spinor inner-product terms neglected in the adiabatic TLM. In this framework, these terms break momentum-space reflection symmetry only for the \((+,-')\) and \((-,+')\) spin configurations, inducing a spin-dependent transverse asymmetry in the momentum distribution. Therefore, the mTLM accurately describes the momentum and phase distributions in multiphoton electron-positron pair production.
	
	For circularly polarized pulses, the momentum distributions exhibit concentric multiphoton rings, while the phase distributions exhibit quantized vortices. The polarization handedness determines the winding direction of the quantized vortices. LCP yields clockwise vortices with \(l_{+,+'} = 2\) and \(l_{-,-'} = 4\), whereas RCP yields counterclockwise vortices with \(l_{+,+'} = -4\) and \(l_{-,-'} = -2\). In contrast, for linearly polarized pulses, the phase distribution exhibits nodal lines instead of quantized vortices, arising from the coherent superposition of equal-amplitude LCP and RCP components.
	
	For two circularly polarized pulses with a time delay, the relative phase drives the vortex evolution through the coherent superposition of the two single-pulse amplitudes. As \(T\) increases, the relative phase varies rapidly with momentum, increasing the number of Ramsey interference rings in the momentum spectrum and that of vortex rings in the phase distribution. For LLCP, constructive interference between the co-rotating pulses doubles the topological charges, yielding \(l_{+,+'} = 4\) and \(l_{-,-'} = 8\), whereas for RLCP, the topological charges are governed by the dominant component in the momentum distribution for each spin configuration, giving \(l_{+,+'} = 2\) and \(l_{-,-'} = -2\). The interference degree \(I_{s,s'}\) quantifies the quantum interference between the two pulses for different spin configurations.
	
	The mTLM provides a self-consistent framework for describing the spin-resolved momentum and phase distributions in multiphoton electron-positron pair production, demonstrating that spin configuration, pulse polarization, time delay, and diagonal couplings jointly govern the vortex structure and its evolution. Future work may extend this analysis to spatially inhomogeneous fields, multicolor fields, finite temperature, and cascade processes.

	\section{Acknowledgments}\label{Sec:V}

		We thank Li-Yi Yang and Zhi-Hang Yao for helpful discussions. This work is supported by the National Natural Science Foundation of China (NSFC) under Grants No. 12375240 and No. 12535015.

	\appendix
	\makeatletter
	\def\section{\@startsection{section}{1}{\z@}%
		{-3.5ex \@plus -1ex \@minus -.2ex}%
		{2.3ex \@plus.2ex}%
		{\normalfont\bfseries\raggedright}}
	\def\@seccntformat#1{\csname the#1\endcsname.\quad}
	\let\@afterindentfalse\@afterindenttrue
	\@afterindenttrue
	\makeatother

	\section{Derivation of the spin-resolved mTLM evolution equations}\label{App:A}

	In the presence of a spatially homogeneous and time-dependent external electric field, the Dirac equation is
	\begin{equation}\label{eq:A1}
		(\mathrm{i}\gamma^{\mu}D_{\mu} - m)\Psi(\bm{x}, t) = 0,
	\end{equation}
	where the covariant derivative is \(D_{\mu} = \partial_{\mu} + \mathrm{i}eA_{\mu}\). The gamma matrices satisfy the anti-commutation relation
	\begin{equation}\label{eq:A2}
		\{\gamma^{\mu},\gamma^{\nu}\} = 2\eta^{\mu \nu}I_{4},
	\end{equation}
	where \(I_{4}\) is the \(4\times 4\) identity matrix. In the Dirac representation, the gamma matrices take the form
	\begin{equation}\label{eq:A3}
		\gamma^0 = \begin{pmatrix} I_2 & O \\ O & -I_2 \end{pmatrix}, \quad \gamma^i = \begin{pmatrix} O & \sigma^i \\ -\sigma^i & O \end{pmatrix},
	\end{equation}
	where \(O\) is the zero matrix, and \(\sigma^{i}\) (\(i = 1,2,3\)) are the Pauli matrices
	\begin{equation}\label{eq:A4}
		\sigma^{1} = \begin{pmatrix} 0 & 1\\ 1 & 0 \end{pmatrix}, \quad \sigma^{2} = \begin{pmatrix} 0 & -\mathrm{i}\\ \mathrm{i} & 0 \end{pmatrix}, \quad \sigma^{3} = \begin{pmatrix} 1 & 0\\ 0 & -1 \end{pmatrix}.
	\end{equation}
	
	Since the external field is spatially homogeneous, the canonical momentum \(\bm{p}\) is a good quantum number, so that the Dirac field operator can be Fourier transformed as
	\begin{equation}\label{eq:A5}
		\Psi (\bm{x},t) = \int \frac{\mathrm{d}^{3}\bm{p}}{(2\pi)^{3}}\psi_{\bm{p}}(t)\mathrm{e}^{\mathrm{i}\bm{p}\cdot \bm{x}},
	\end{equation}
	where \(\psi_{\bm{p}}(t)\) is the momentum-space spinor. Substituting Eq.~\eqref{eq:A5} into Eq.~\eqref{eq:A1}, we obtain the momentum-space Dirac equation
	\begin{equation}\label{eq:A6}
		\mathrm{i}\partial_{t}\psi_{\bm{p}}(t) = H_{\bm{p}}(t)\psi_{\bm{p}}(t),
	\end{equation}
	where the Dirac Hamiltonian is \(H_{\bm{p}}(t) = \gamma^{0}\bm{\gamma}\cdot \bm{P}(t) + \gamma^{0}m\), with the kinetic momentum defined as \(\bm{P}(t) = \bm{p} - e\bm{A}(t)\). The instantaneous eigenstates \(u_{\bm{p},s}(t)\) and \(v_{-\bm{p},s'}(t)\) satisfy
	\begin{subequations}\label{eq:A7}
		\begin{align}
			H_{\bm{p}}(t)u_{\bm{p},s}(t) &= +\omega_{\bm{p}}(t)u_{\bm{p},s}(t), \label{eq:A7a} \\
			H_{\bm{p}}(t)v_{-\bm{p},s'}(t) &= -\omega_{\bm{p}}(t)v_{-\bm{p},s'}(t), \label{eq:A7b}
		\end{align}
	\end{subequations}
	where the instantaneous eigenvalues are \(\pm \omega_{\bm{p}}(t) = \pm \sqrt{\bm{P}^{2}(t) + m^{2}}\). Expanding the Dirac field operator in terms of the instantaneous eigenstates yields
	\begin{equation}\label{eq:A8}
		\Psi (\bm{x},t) = \sum_{s = \pm 1}\sum_{s' = \pm 1}\int \frac{\mathrm{d}^{3}\bm{p}}{(2\pi)^{3}} [a_{\bm{p},s}u_{\bm{p},s}(t) + b_{-\bm{p},s'}^{\dagger}v_{-\bm{p},s'}(t)]\mathrm{e}^{\mathrm{i}\bm{p}\cdot \bm{x}},
	\end{equation}
	where \(a_{\bm{p},s}\) and \(b_{-\bm{p},s'}^{\dagger}\) are the electron annihilation and positron creation operators, respectively, with \(s\) and \(s'\) as the spin projections of the electron and positron along the \(z\)-axis.
	
	Solving the eigenvalue equations Eq.~\eqref{eq:A7} with the normalization conditions \(u_{\bm{p},s}^{\dagger}(t)u_{\bm{p},s}(t) = 1\) and \(v_{-\bm{p},s'}^{\dagger}(t)v_{-\bm{p},s'}(t) = 1\) yields the normalized spinor eigenstates:
	\begin{subequations}\label{eq:A9}
		\begin{equation}\label{eq:A9a}
			u_{\bm{p},+1}(t) = \frac{1}{\sqrt{2\omega_{\bm{p}}(t)[\omega_{\bm{p}}(t) + m]}}\begin{pmatrix} \omega_{\bm{p}}(t) + m \\ 0 \\ P_z(t) \\ P_x(t) + \mathrm{i}P_y(t) \end{pmatrix},
		\end{equation}
		
		\begin{equation}\label{eq:A9b}
			u_{\bm{p},-1}(t) = \frac{1}{\sqrt{2\omega_{\bm{p}}(t)[\omega_{\bm{p}}(t) + m]}}\begin{pmatrix} 0 \\ \omega_{\bm{p}}(t) + m \\ P_x(t) - \mathrm{i}P_y(t) \\ -P_z(t) \end{pmatrix},
		\end{equation}
		
		\begin{equation}\label{eq:A9c}
			v_{-\bm{p},+1}(t) = \frac{1}{\sqrt{2\omega_{\bm{p}}(t)[\omega_{\bm{p}}(t) + m]}}\begin{pmatrix} -P_x(t) + \mathrm{i}P_y(t) \\ P_z(t) \\ 0 \\ \omega_{\bm{p}}(t) + m \end{pmatrix},
		\end{equation}
		
		\begin{equation}\label{eq:A9d}
			v_{-\bm{p},-1}(t) = \frac{1}{\sqrt{2\omega_{\bm{p}}(t)[\omega_{\bm{p}}(t) + m]}}\begin{pmatrix} -P_z(t) \\ -P_x(t) - \mathrm{i}P_y(t) \\ \omega_{\bm{p}}(t) + m \\ 0 \end{pmatrix}.
		\end{equation}
	\end{subequations}
	These eigenstates satisfy the orthonormality relations \(u_{\bm{p},s}^{\dagger}(t)u_{\bm{p},s'}(t) = v_{-\bm{p},s}^{\dagger}(t)v_{-\bm{p},s'}(t) = \delta_{s,s'}\) and \(u_{\bm{p},s}^{\dagger}(t)v_{-\bm{p},s'}(t) = v_{-\bm{p},s'}^{\dagger}(t)u_{\bm{p},s}(t) = 0\), and form a complete basis at each instant \(t\).
	
	Using the explicit spinor forms in Eqs.~\eqref{eq:A9a}--\eqref{eq:A9d}, we obtain the inner-product relations \(u_{\bm{p},s}^{\dagger}(t)\dot{u}_{\bm{p},s'}(t) = v_{-\bm{p},-s}^{\dagger}(t)\dot{v}_{-\bm{p},-s'}(t)\) and \(u_{\bm{p},s}^{\dagger}(t)\dot{v}_{-\bm{p},s'}(t) = -[v_{-\bm{p},s}^{\dagger}(t)\dot{u}_{\bm{p},s'}(t)]^*\). Accordingly, we define the spinor inner products as \(\Delta_{s,s'}(\bm{p},t) \equiv -\mathrm{i}u_{\bm{p},s}^{\dagger}(t)\dot{u}_{\bm{p},s'}(t)\) and \(\Omega_{s,s'}(\bm{p},t) \equiv -u_{\bm{p},s}^{\dagger}(t)\dot{v}_{-\bm{p},s'}(t)\), which take the form
	\begin{equation}\label{eq:A10}
		\begin{aligned}
			\Delta_{+,+'}(\bm{p},t) &= +\mu(t)\mathcal{F}_z(t), \\
			\Delta_{+,-'}(\bm{p},t) &= \mu(t)\mathcal{F}_x(t) - \mathrm{i}\mu(t)\mathcal{F}_y(t), \\
			\Delta_{-,+'}(\bm{p},t) &= \mu(t)\mathcal{F}_x(t) + \mathrm{i}\mu(t)\mathcal{F}_y(t), \\
			\Delta_{-,-'}(\bm{p},t) &= -\mu(t)\mathcal{F}_z(t),
		\end{aligned}
	\end{equation}
	and
	\begin{equation}\label{eq:A11}
		\begin{aligned}
			\Omega_{+,+'}(\bm{p},t) &= v_1(t)[E_x(t) - \mathrm{i}E_y(t)] - v_2(t)[P_x(t) - \mathrm{i}P_y(t)], \\
			\Omega_{+,-'}(\bm{p},t) &= +v_1(t)E_z(t) - v_2(t)P_z(t), \\
			\Omega_{-,+'}(\bm{p},t) &= -v_1(t)E_z(t) + v_2(t)P_z(t), \\
			\Omega_{-,-'}(\bm{p},t) &= v_1(t)[E_x(t) + \mathrm{i}E_y(t)] - v_2(t)[P_x(t) + \mathrm{i}P_y(t)],
		\end{aligned}
	\end{equation}
	where the auxiliary quantities are defined as \(\bm{\mathcal{F}}(t) = \bm{P}(t) \times \bm{E}(t)\), \(\mu(t) = \frac{e}{2\omega_{\bm{p}}(t)[\omega_{\bm{p}}(t) + m]}\), \(v_1(t) = \frac{e}{2\omega_{\bm{p}}(t)}\), and \(v_2(t) = \frac{e\bm{P}(t)\cdot\bm{E}(t)}{2\omega_{\bm{p}}^2(t)[\omega_{\bm{p}}(t) + m]}\).
	
	Expanding the momentum-space spinor in the instantaneous eigenbasis yields
	\begin{equation}\label{eq:A12}
		\psi_{s,s'}(\bm{p},t) = \alpha_{s,s'}(\bm{p},t)u_{\bm{p},s}(t)\mathrm{e}^{-\mathrm{i}\int_{-\infty}^{t}\omega_{\bm{p}}(\tau)\mathrm{d}\tau} + \beta_{s,s'}(\bm{p},t)v_{-\bm{p},s'}(t)\mathrm{e}^{+\mathrm{i}\int_{-\infty}^{t}\omega_{\bm{p}}(\tau)\mathrm{d}\tau},
	\end{equation}
	where \(\alpha_{s,s'}(\bm{p},t)\) and \(\beta_{s,s'}(\bm{p},t)\) are the Bogoliubov coefficients, and \(\mathrm{e}^{\mp\mathrm{i}\int_{-\infty}^{t}\omega_{\bm{p}}(\tau)\mathrm{d}\tau}\) are the corresponding dynamical phase factors.
	
	By substituting Eq.~\eqref{eq:A12} into the momentum-space Dirac equation \eqref{eq:A6} and using the instantaneous eigenvalue equation \eqref{eq:A7} together with the orthonormality relations of the instantaneous eigenstates, we obtain the coupled evolution equations for the Bogoliubov coefficients:
	\begin{equation}\label{eq:A13}
		\begin{aligned}
			\dot{\alpha}_{s,s'}(\bm{p},t) &= -\mathrm{i}\Delta_{s,s'}(\bm{p},t)\alpha_{s,s'}(\bm{p},t) + \Omega_{s,s'}(\bm{p},t)\beta_{s,s'}(\bm{p},t)\mathrm{e}^{+2\mathrm{i}\int_{-\infty}^{t}\omega_{\bm{p}}(\tau)\mathrm{d}\tau}, \\
			\dot{\beta}_{s,s'}(\bm{p},t) &= -\mathrm{i}\Delta_{-s,-s'}(\bm{p},t)\beta_{s,s'}(\bm{p},t) - \Omega_{s,s'}^{*}(\bm{p},t)\alpha_{s,s'}(\bm{p},t)\mathrm{e}^{-2\mathrm{i}\int_{-\infty}^{t}\omega_{\bm{p}}(\tau)\mathrm{d}\tau}.
		\end{aligned}
	\end{equation}
	
	To absorb the fast-oscillating dynamical phase factors, we introduce the phase-rotated coefficients
	\begin{equation}\label{eq:A14}
		\begin{aligned}
			c_{s,s'}^{\alpha}(\bm{p},t) &= \alpha_{s,s'}(\bm{p},t)\mathrm{e}^{-\mathrm{i}\int_{-\infty}^{t}\omega_{\bm{p}}(\tau)\mathrm{d}\tau}, \\
			c_{s,s'}^{\beta}(\bm{p},t) &= \beta_{s,s'}(\bm{p},t)\mathrm{e}^{+\mathrm{i}\int_{-\infty}^{t}\omega_{\bm{p}}(\tau)\mathrm{d}\tau}.
		\end{aligned}
	\end{equation}
	
	Taking the partial time derivative of Eq.~\eqref{eq:A14} and substituting Eq.~\eqref{eq:A13} yields the two-level evolution equations:
	\begin{equation}\label{eq:A15}
		\begin{aligned}
			\dot{c}_{s,s'}^{\alpha}(\bm{p},t) &= -\mathrm{i}[\omega_{\bm{p}}(t) + \Delta_{s,s'}(\bm{p},t)]c_{s,s'}^{\alpha}(\bm{p},t) + \Omega_{s,s'}(\bm{p},t)c_{s,s'}^{\beta}(\bm{p},t), \\
			\dot{c}_{s,s'}^{\beta}(\bm{p},t) &= +\mathrm{i}[\omega_{\bm{p}}(t) - \Delta_{-s,-s'}(\bm{p},t)]c_{s,s'}^{\beta}(\bm{p},t) - \Omega_{s,s'}^{*}(\bm{p},t)c_{s,s'}^{\alpha}(\bm{p},t).
		\end{aligned}
	\end{equation}
	
	Casting Eq.~\eqref{eq:A15} into matrix form yields the spin-resolved two-level evolution equation:
	\begin{equation}\label{eq:A16}
		\mathrm{i} \begin{pmatrix} \dot{c}_{s,s'}^{\alpha}(\bm{p},t) \\ \dot{c}_{s,s'}^{\beta}(\bm{p},t) \end{pmatrix} = \begin{pmatrix} \omega_{\bm{p}}(t) + \Delta_{s,s'}(\bm{p},t) & \mathrm{i}\Omega_{s,s'}(\bm{p},t) \\ -\mathrm{i}\Omega_{s,s'}^{*}(\bm{p},t) & -\omega_{\bm{p}}(t) + \Delta_{-s,-s'}(\bm{p},t) \end{pmatrix} \begin{pmatrix} c_{s,s'}^{\alpha}(\bm{p},t) \\ c_{s,s'}^{\beta}(\bm{p},t) \end{pmatrix},
	\end{equation}
	which is exactly the spin-resolved evolution equation of the mTLM, as given by Eq.~\eqref{eq:2} in Sec.~\ref{Sec:II}.

	\section{Time-delay phase factor from temporal translation}\label{App:B}

	With the two-level state vector defined as \(|\phi_{\bm{p}}(t)\rangle = (c_{s,s'}^{\alpha}(\bm{p},t), c_{s,s'}^{\beta}(\bm{p},t))^{T}\), we decompose the effective Hamiltonian into a free part \(\mathcal{H}_{0}(\bm{p},t)\) and an interaction part \(\mathcal{H}(\bm{p},t)\), which rewrites the mTLM evolution equation as
	\begin{equation}\label{eq:B1}
		\mathrm{i}\frac{\mathrm{d}}{\mathrm{d}t} |\phi_{\bm{p}}(t)\rangle = [\mathcal{H}_{0}(\bm{p},t) + \mathcal{H}(\bm{p},t)]|\phi_{\bm{p}}(t)\rangle,
	\end{equation}
	where \(\mathcal{H}_{0}(\bm{p},t)\) and \(\mathcal{H}(\bm{p},t)\) govern the fast intrinsic oscillations and slow field-driven transitions, respectively, taking the form
	\begin{equation}\label{eq:B2}
		\mathcal{H}_{0}(\bm{p},t) = \begin{pmatrix} \omega_{\bm{p}}(t) & 0\\ 0 & -\omega_{\bm{p}}(t) \end{pmatrix}, \quad \mathcal{H}(\bm{p},t) = \begin{pmatrix} \Delta_{s,s'}(\bm{p},t) & \mathrm{i}\Omega_{s,s'}(\bm{p},t)\\ -\mathrm{i}\Omega_{s,s'}^{*}(\bm{p},t) & \Delta_{-s,-s'}(\bm{p},t) \end{pmatrix}.
	\end{equation}
	
	To separate these two time scales, we transform from the Schrödinger picture to the interaction picture. The free evolution operator is
	\begin{equation}\label{eq:B3}
		M_{0}(\bm{p},t) = \mathrm{e}^{-\mathrm{i}\int_{-\infty}^{t}\mathcal{H}_{0}(\bm{p},\tau)\mathrm{d}\tau}.
	\end{equation}
	
	The Schrödinger-picture propagator \(W_{0}(\bm{p},t)\) and the interaction-picture propagator \(W(\bm{p},t)\) are related by \(W_{0}(\bm{p},t) = M_{0}(\bm{p},t)W(\bm{p},t)\). Given the initial condition \(W_{0}(\bm{p},-\infty) = I\), the state vector evolves as
	\begin{equation}\label{eq:B4}
		|\phi_{\bm{p}}(t)\rangle = W_{0}(\bm{p},t)|\phi_{\bm{p}}(-\infty)\rangle.
	\end{equation}
	
	For the initial vacuum state \(|\phi_{\bm{p}}(-\infty)\rangle = (1,0)^{T}\), the probability amplitude \(c_{s,s'}^{\beta}(\bm{p},t)\) is exactly the (2,1) matrix element of the propagator:
	\begin{equation}\label{eq:B5}
		c_{s,s'}^{\beta}(\bm{p},t) = [W_{0}(\bm{p},t)]_{21}.
	\end{equation}
	
	To investigate the effect of the relative time delay \(T\), we consider a temporal translation of the system. Correspondingly, we introduce \(T\) as an additional parameter separated by a semicolon in the arguments of all relevant quantities, with \(T = 0\) as the zero-delay reference. The temporal translation ensures that the interaction Hamiltonian satisfies \(\mathcal{H}(\bm{p},t;T) = \mathcal{H}(\bm{p},t - T;0)\), which in turn dictates that the interaction-picture propagator satisfies \(W(\bm{p},t;T) = W(\bm{p},t - T;0)\). Transforming back to the Schrödinger picture yields \(W_{0}(\bm{p},t;T) = M_{0}(\bm{p},t)M_{0}^{-1}(\bm{p},t - T;0)W_{0}(\bm{p},t - T;0)\), whose (2,1) matrix element is
	\begin{equation}\label{eq:B6}
		[W_{0}(\bm{p},t;T)]_{21} = \mathrm{e}^{\mathrm{i}\int_{t - T}^{t}\omega_{\bm{p}}(\tau)\mathrm{d}\tau}[W_{0}(\bm{p},t - T;0)]_{21}.
	\end{equation}
	
	For \(T\) much shorter than the characteristic time scale of the instantaneous energy variation, the integral is approximated as \(\int_{t - T}^{t}\omega_{\bm{p}}(\tau)\mathrm{d}\tau \approx \omega_{\bm{p}}(t)T\). At the final time \(t\rightarrow +\infty\), the propagator reaches its asymptotic value, so that \([W_{0}(\bm{p},+\infty - T;0)]_{21} = [W_{0}(\bm{p},+\infty;0)]_{21}\). Consequently, the probability amplitude for the delayed pulse acquires a phase factor relative to the zero-delay reference:
	\begin{equation}\label{eq:B7}
		c_{s,s'}^{\beta}(\bm{p},+\infty;T) = \mathrm{e}^{\mathrm{i}\omega_{\bm{p}}(+\infty)T}c_{s,s'}^{\beta}(\bm{p},+\infty;0).
	\end{equation}
	This temporal translation represents the relative time delay of the second pulse with respect to the first. Consequently, Eq.~\eqref{eq:B7} directly yields the delayed single-pulse probability amplitude \(c_{s,s'}^{\beta,(2)}(\bm{p})\mathrm{e}^{\mathrm{i}\omega_{\bm{p}}T}\), which forms the second term of Eq.~\eqref{eq:8}.

	\section{Momentum reflection transformation}\label{App:C}

	The evolution equations for the mTLM are
	\begin{equation}\label{eq:C1}
		\begin{aligned}
			\dot{c}_{s,s'}^{\alpha}(\bm{p},t) &= -\mathcal{U}_{s,s'}(\bm{p},t)c_{s,s'}^{\alpha}(\bm{p},t) + \Omega_{s,s'}(\bm{p},t)c_{s,s'}^{\beta}(\bm{p},t), \\
			\dot{c}_{s,s'}^{\beta}(\bm{p},t) &= +\mathcal{V}_{-s,-s'}(\bm{p},t)c_{s,s'}^{\beta}(\bm{p},t) - \Omega_{s,s'}^{*}(\bm{p},t)c_{s,s'}^{\alpha}(\bm{p},t),
		\end{aligned}
	\end{equation}
	where \(\mathcal{U}_{s,s'}(\bm{p},t) = \mathrm{i}[\omega_{\bm{p}}(t) + \Delta_{s,s'}(\bm{p},t)]\) and \(\mathcal{V}_{-s,-s'}(\bm{p},t) = \mathrm{i}[\omega_{\bm{p}}(t) - \Delta_{-s,-s'}(\bm{p},t)]\).
	
	Applying the momentum reflection transformation \(\mathcal{P}\colon \bm{p} \to \bm{p}^{\mathcal{P}}\) to Eq.~\eqref{eq:C1} yields
	\begin{equation}\label{eq:C2}
		\begin{aligned}
			\dot{c}_{s,s'}^{\alpha}(\bm{p}^{\mathcal{P}},t) &= -\mathcal{U}_{s,s'}(\bm{p}^{\mathcal{P}},t)c_{s,s'}^{\alpha}(\bm{p}^{\mathcal{P}},t) + \Omega_{s,s'}(\bm{p}^{\mathcal{P}},t)c_{s,s'}^{\beta}(\bm{p}^{\mathcal{P}},t), \\
			\dot{c}_{s,s'}^{\beta}(\bm{p}^{\mathcal{P}},t) &= +\mathcal{V}_{-s,-s'}(\bm{p}^{\mathcal{P}},t)c_{s,s'}^{\beta}(\bm{p}^{\mathcal{P}},t) - \Omega_{s,s'}^{*}(\bm{p}^{\mathcal{P}},t)c_{s,s'}^{\alpha}(\bm{p}^{\mathcal{P}},t).
		\end{aligned}
	\end{equation}
	
	Introducing the linear transformation
	\begin{subequations}\label{eq:C3}
		\begin{equation}\label{eq:C3a}
			c_{s,s'}^{\alpha}(\bm{p}^{\mathcal{P}},t) = \mathcal{A}(t)c_{s,s'}^{\alpha}(\bm{p},t),
		\end{equation}
		\begin{equation}\label{eq:C3b}
			c_{s,s'}^{\beta}(\bm{p}^{\mathcal{P}},t) = \mathcal{B}(t)c_{s,s'}^{\beta}(\bm{p},t).
		\end{equation}
	\end{subequations}
	
	Differentiating Eqs.~\eqref{eq:C3a} and \eqref{eq:C3b} with respect to time, we obtain
	\begin{equation}\label{eq:C4}
		\begin{aligned}
			\dot{c}_{s,s'}^{\alpha}(\bm{p}^{\mathcal{P}},t) &= \dot{\mathcal{A}}(t)c_{s,s'}^{\alpha}(\bm{p},t) + \mathcal{A}(t)\dot{c}_{s,s'}^{\alpha}(\bm{p},t), \\
			\dot{c}_{s,s'}^{\beta}(\bm{p}^{\mathcal{P}},t) &= \dot{\mathcal{B}}(t)c_{s,s'}^{\beta}(\bm{p},t) + \mathcal{B}(t)\dot{c}_{s,s'}^{\beta}(\bm{p},t).
		\end{aligned}
	\end{equation}
	
	Upon substituting \textcolor{blue}{Eq. (C1)} into \textcolor{blue}{Eq. (C4)} and \textcolor{blue}{Eqs. (C3a) and (C3b)} into \textcolor{blue}{Eq. (C2)}, equating the respective coefficients of \(c_{s,s'}^{\alpha}(\boldsymbol{p},t)\) and \(c_{s,s'}^{\beta}(\boldsymbol{p},t)\) yields
	\begin{subequations}\label{eq:C5}
		\begin{align}
			\dot{\mathcal{A}}(t) &= \mathcal{A}(t)[\mathcal{U}_{s,s'}(\bm{p},t) - \mathcal{U}_{s,s'}(\bm{p}^{\mathcal{P}},t)], \label{eq:C5a} \\
			\dot{\mathcal{B}}(t) &= \mathcal{B}(t)[\mathcal{V}_{-s,-s'}(\bm{p}^{\mathcal{P}},t) - \mathcal{V}_{-s,-s'}(\bm{p},t)], \label{eq:C5b} \\
			\mathcal{A}(t) &= \frac{\Omega_{s,s'}^{*}(\bm{p},t)}{\Omega_{s,s'}^{*}(\bm{p}^{\mathcal{P}},t)}\mathcal{B}(t), \label{eq:C5c} \\
			\mathcal{B}(t) &= \frac{\Omega_{s,s'}(\bm{p},t)}{\Omega_{s,s'}(\bm{p}^{\mathcal{P}},t)}\mathcal{A}(t). \label{eq:C5d}
		\end{align}
	\end{subequations}
	
	Substituting Eq.~\eqref{eq:C5c} into Eq.~\eqref{eq:C5d} yields the constraint
	\begin{equation}\label{eq:C6}
		\frac{\Omega_{s,s'}(\bm{p},t)}{\Omega_{s,s'}(\bm{p}^{\mathcal{P}},t)}\frac{\Omega_{s,s'}^{*}(\bm{p},t)}{\Omega_{s,s'}^{*}(\bm{p}^{\mathcal{P}},t)} = \left|\frac{\Omega_{s,s'}(\bm{p},t)}{\Omega_{s,s'}(\bm{p}^{\mathcal{P}},t)}\right|^{2} = 1.
	\end{equation}
	
	With the initial amplitude \(c_{s,s'}^{\alpha}(\bm{p}, -\infty) = c_{s,s'}^{\alpha}(\bm{p}^{\mathcal{P}}, -\infty) = 1\), Eq.~\eqref{eq:C3a} directly yields \(\mathcal{A}(-\infty) = 1\). Integrating Eq.~\eqref{eq:C5a} for \(\mathcal{A}(t)\) subsequently yields
	\begin{equation}\label{eq:C7}
		\mathcal{A}(t) = \exp \left\{\int_{-\infty}^{t}[\mathcal{U}_{s,s'}(\bm{p},\tau) - \mathcal{U}_{s,s'}(\bm{p}^{\mathcal{P}},\tau)]\mathrm{d}\tau \right\}.
	\end{equation}
	
	Combining Eqs.~\eqref{eq:C3b}, \eqref{eq:C5d}, and \eqref{eq:C7}, the amplitude \(c_{s,s'}^{\beta}(\bm{p}, t)\) transforms under momentum reflection as
	\begin{equation}\label{eq:C8}
		\frac{c_{s,s'}^{\beta}(\bm{p}^{\mathcal{P}},t)}{c_{s,s'}^{\beta}(\bm{p},t)} = \frac{\Omega_{s,s'}(\bm{p},t)}{\Omega_{s,s'}(\bm{p}^{\mathcal{P}},t)}\exp \left\{\int_{-\infty}^{t}[\mathcal{U}_{s,s'}(\bm{p},\tau) - \mathcal{U}_{s,s'}(\bm{p}^{\mathcal{P}},\tau)]\mathrm{d}\tau \right\}.
	\end{equation}
	
	Taking the squared modulus of Eq.~\eqref{eq:C8} and applying Eq.~\eqref{eq:C6}, the ratio of the momentum distributions at \(t \to +\infty\) reduces to
	\begin{equation}\label{eq:C9}
		\frac{f_{s,s'}(\bm{p}^{\mathcal{P}})}{f_{s,s'}(\bm{p})} = \exp \left\{2\int_{-\infty}^{+\infty}\mathrm{Im}\left[\Delta_{s,s'}(\bm{p}^{\mathcal{P}},t) - \Delta_{s,s'}(\bm{p},t)\right]\mathrm{d}t\right\},
	\end{equation}
	which mathematically reveals that the transverse asymmetry originates entirely from the imaginary part of the diagonal spinor inner-product terms. The physical consequences for different spin configurations are analyzed in Sec.~\ref{SecIII:C}.

\end{document}